\documentclass[aps,prd,nofootinbib, twocolumn, superscriptaddress]{revtex4-2}

\usepackage{amsmath,amssymb, amsthm,amstext}
\usepackage{natbib}
\usepackage{graphicx}
\usepackage{color}
\usepackage{array, enumerate}
\usepackage{bm}
\usepackage{multirow}
\usepackage[breaklinks,colorlinks,citecolor=blue]{hyperref}
\usepackage{braket}
\usepackage{txfonts}
\usepackage{physics}
\usepackage{xcolor}

\usepackage{soul}
\setulcolor{magenta}
\setul{-2.5pt}{1.5pt}

\def\be{\begin{equation}}
\def\ee{\end{equation}}
\def\bw{\begin{widetext}}
\def\ew{\end{widetext}}
\def\ba{\begin{aligned}}
\def\ea{\end{aligned}}
\def\bpm{\begin{pmatrix}}
\def\epm{\end{pmatrix}}

\begin{document}

\title{Probing superradiant cloud level transitions in stellar-mass black hole binaries with milli-Hertz and multiband gravitational wave observations}
\author{Si-Tong Peng}
\email{pengsitong21@mails.ucas.ac.cn}
\affiliation{International Centre for Theoretical Physics Asia-Pacific, University of Chinese Academy of Sciences, 100190 Beijing, China}
\author{Anson Chen}
\email{chena@ucas.ac.cn}
\affiliation{International Centre for Theoretical Physics Asia-Pacific, University of Chinese Academy of Sciences, 100190 Beijing, China}
\author{Si-Yang Pu}
\email{pusiyang25@mails.ucas.ac.cn}
\affiliation{International Centre for Theoretical Physics Asia-Pacific, University of Chinese Academy of Sciences, 100190 Beijing, China}
\author{Jun Zhang}
\email{zhangjun@ucas.ac.cn}
\affiliation{International Centre for Theoretical Physics Asia-Pacific, University of Chinese Academy of Sciences, 100190 Beijing, China}
\affiliation{Taiji Laboratory for Gravitational Wave Universe (Beijing/Hangzhou), University of Chinese Academy of Sciences, 100049 Beijing, China}

\begin{abstract}

We investigate superradiant cloud level transitions in stellar-mass black hole binaries with milli-Hertz and multiband gravitational-wave observations. Under the adopted cloud formation and survival conditions, the $\lvert322\rangle\rightarrow\lvert31-1\rangle$ fine transition can imprint phase corrections in the LISA band, while Bohr transitions occurring after the milli-Hertz band observation can be probed through shifts in the merger time. Modeling the coupled cloud-orbit evolution, we find that cloud depletion can begin before resonance and that ionization can advance rather than delay coalescence. For a representative nearby binary, a vacuum injection-recovery analysis yields a cloud-mass-fraction posterior concentrated at $M_c^i/M\lesssim10^{-6}$. Comparing the LISA-predicted coalescence time with a subsequent ground-based measurement provides a complementary test of later cloud evolution. These results motivate complementary searches for ultralight bosons with milli-Hertz and multiband gravitational-wave observations.

\end{abstract}
\maketitle

\section{Introduction}\label{sec:introduction}

Ultralight bosonic fields, such as axion-like particles, are motived in many beyond Standard Model theoreies~\cite{Arvanitaki:2009fg} and are proposed as possible dark matter candidates~\cite{Hui:2016ltb}. If the Compton wavelength of a boson field is comparable to the size of a rapidly rotating black hole, the field can experience superradiant growth, forming a cloud around the black hole~\cite{Brito:2015oca, Arvanitaki:2014wva}. The resulting system, often referred to as a gravitational atom due to its structural similarity to a hydrogen atom, provides a promising laboratory for probing ultralight bosons~\cite{Brito:2017wnc, Brito:2017zvb, Tsukada:2018mbp, Isi:2018pzk, Palomba:2019vxe,Tsukada:2020lgt,Ng:2020ruv,LIGOScientific:2021rnv,Yuan:2022bem, Aurrekoetxea:2023jwk, Aurrekoetxea:2024cqd, Miller:2025yyx}. 

Superradiant clouds can also form around black holes in binary systems, where they can affect the orbital dynamics and imprint signatures on the gravitational waves (GWs) emitted during binary coalescences, thereby providing an additional approach for probing ultralight bosons with GW observations. In particular, a cloud initially occupying one eigenstate can undergo level transitions~\cite{Baumann:2018vus,Baumann:2019ztm,Berti:2019wnn}, ionization~\cite{Baumann:2021fkf, Guo:2025pea}, mass transfer~\cite{Guo:2023lbv,Guo:2024iye}, and even form a common envelope~\cite{Guo:2025ckp}. Also see Refs~\cite{Aurrekoetxea:2023jwk, Aurrekoetxea:2024cqd, Zhang:2026buv} for other interesting phenomena of ultralight boson fields in binaries. These processes generally involve exchanges of energy and angular momentum between the superradiant cloud and the binary orbit, and can therefore modify the orbital decay rate~\cite{Cardoso:2011xi, Ferreira:2017pth, Zhang:2018kib, Zhang:2019eid, Baumann:2022pkl, Tomaselli:2023ysb, Cao:2023fyv, Tomaselli:2024bdd, Cao:2024wby, Takahashi:2024fyq, Zhu:2024bqs, Guo:2023lbv, Guo:2025ckp} as well as the evolution of orbital parameters such as the eccentricity and inclination~\cite{Tomaselli:2024bdd,Tomaselli:2024dbw,Boskovic:2024fga,Guo:2025ckp}. These processes could also be accompanied by GW radiation \cite{Kyriazis:2025fis, Peng:2025zca}.

Stellar-mass black hole binaries (SMBHBs) provide a setting in which these dynamical signatures can be tested across widely separated GW frequency bands. Clouds in SMBHBs are generally expected to be depleted before the binaries enter the detector band, and hence one does not generically expect to detect a live boson cloud during the observed late inspiral of LVK BBHs\footnote{However, see Ref.~\cite{Roy:2025qaa} for hints of a scalar environment in GW190728 and GW190814.}. Nevertheless, earlier cloud-binary interactions may leave indirect evidence. In particular, residual eccentricity could survive as a fossil signature of a boson cloud, even if the cloud itself has already been consumed before observation~\cite{Tomaselli:2024dbw,Boskovic:2024fga}. On the other hand, the planned space based GW detectors, such as LISA, Taiji and Tianqin, can observe the early inspiral of sufficiently nearby SMBHBs in the milli-Hertz band, years to decades before their mergers become accessible to ground-based detectors. With such detectors, we may observe the fingerprints of cloud level transition directly. In this work, we investigate the probes of scalar cloud level transitions in SMBHBs, using LISA as the representative milli-Hertz detector. We will also show that even if the transition happens between the milli-Hertz and kilohertz bands, its effects on the merger time can be inferred with multiband observations~\cite{Chen:2024ery}.

The paper is organized as follows. Section~\ref{SMBHB} describes the SMBHB population and the selection criteria for milli-Hertz and multiband observations. Section~\ref{sec:signatures} identifies the relevant transitions and computes their effects on the GW phase and merger time. Section~\ref{sec:detectability} presents the parameter-inference benchmarks and multiband timing test. We conclude in Sec.~\ref{sec:conclusion}. We use geometric units, $G=c=\hbar=1$.

\section{Stellar-mass black hole binaries in the milli-Hertz band}\label{SMBHB}

We first estimate the population of SMBHBs accessible to space-based GW observations and to multiband observations combining space- and ground-based detectors. These estimates provide the source population relevant to searches for cloud-induced orbital effects. We specify the underlying binary population, construct the selection criteria for the two observing scenarios, and evaluate the resulting detection counts and source-parameter distributions. The forecasts in this section use quasicircular vacuum-binary waveforms; they do not yet impose any requirements on the formation or detectability of boson clouds. 

\subsection{Source population}\label{sec:source_population}

We describe the intrinsic SMBHB population by a normalized source-frame mass distribution, $p(m_1,m_2)$, and a merger-rate density, ${\cal R}(z)$, per unit comoving volume and source-frame time. We order the component masses as $m_1\geq m_2$ and define the mass ratio as $q=m_2/m_1\leq1$. For this forecast, we neglect redshift evolution of the mass distribution and write $p(m_1,m_2)=p(m_1)p(m_2|m_1)$.

For the primary mass, we adopt the Power Law + Multi-Peak prescription considered in the GWTC-4 analysis of Ref.~\cite{LIGOScientific:2025jau}. Before applying the low-mass taper, the distribution is a mixture of a power law and two Gaussian components,
\begin{align}
    p_0(m_1) ={}& (1-\lambda_{\rm g}){\cal P}(m_1|M_{\rm min},M_{\rm max},\alpha_p) \nonumber\\
    &+\lambda_{\rm g}\lambda_{\rm g,low}{\cal G}(m_1|\mu_{\rm g,low},\sigma_{\rm g,low}) \nonumber\\
    &+\lambda_{\rm g}(1-\lambda_{\rm g,low}){\cal G}(m_1|\mu_{\rm g,high},\sigma_{\rm g,high}).
    \label{eq:mass_model}
\end{align}
Here ${\cal P}\propto m_1^{-\alpha_p}$ is a single truncated power law, and ${\cal G}$ denotes a truncated Gaussian; all three components are normalized on $[M_{\rm min},M_{\rm max}]$. The coefficients $\lambda_{\rm g}$ and $\lambda_{\rm g,low}$ specify the total Gaussian weight and the relative weight of the lower-mass peak before tapering. The two peaks have means $\mu_{\rm g,low}$ and $\mu_{\rm g,high}$ and widths $\sigma_{\rm g,low}$ and $\sigma_{\rm g,high}$, respectively. Following Ref.~\cite{LIGOScientific:2025jau}, we apply a smooth low-mass cutoff, $S(m|M_{\rm min},\delta_m)$, which rises from zero to unity over $M_{\rm min}<m<M_{\rm min}+\delta_m$. The normalized primary-mass distribution is then $p(m_1)\propto p_0(m_1)S(m_1|M_{\rm min},\delta_m)$. The conditional secondary-mass distribution is proportional to $q^{\beta_p}S(m_2|M_{\rm min},\delta_m)$ and is normalized over $M_{\rm min}\leq m_2\leq m_1$ at fixed $m_1$.

We describe the redshift dependence of the merger-rate density using the Madau-Dickinson parametrization~\cite{Madau:2014bja,LIGOScientific:2025jau},
\begin{equation}
  {\cal R}(z) = R_0 \left[1+(1+z_p)^{-\gamma-\kappa}\right]
  \frac{(1+z)^\gamma}
  {1+\left(\frac{1+z}{1+z_p}\right)^{\gamma+\kappa}}.
  \label{eq:merger_rate}
\end{equation}
The prefactor ensures ${\cal R}(0)=R_0$, where $R_0$ is the local merger-rate density. The indices $\gamma$ and $\kappa$ control the low- and high-redshift scalings, ${\cal R}\propto(1+z)^\gamma$ and ${\cal R}\propto(1+z)^{-\kappa}$, while $z_p$ sets the characteristic turnover scale. We use this form as a phenomenological description of merger-rate evolution, rather than assuming that the merger rate follows the star-formation rate exactly. The fiducial mass and redshift hyperparameters are listed in Table~\ref{tab:GWTC4_param}; the rate normalization is specified in Sec.~\ref{sec:detection_forecasts}.

\begin{table}
    \centering
    \caption{Fiducial hyperparameters adopted for the Power Law + Multi-Peak mass model and the Madau-Dickinson merger-rate model. Mass parameters are given in the source frame.}
    \label{tab:GWTC4_param}
    \begin{ruledtabular}
    \begin{tabular}{lrlr}
        Parameter & Value & Parameter & Value \\
        \hline
        $\alpha_p$ & $2.9$ & $\mu_{\rm g,low}\,[M_\odot]$ & $9.7$ \\
        $\beta_p$ & $1.0$ & $\mu_{\rm g,high}\,[M_\odot]$ & $30.7$ \\
        $M_{\rm min}\,[M_\odot]$ & $4.6$ & $\sigma_{\rm g,low}\,[M_\odot]$ & $0.7$ \\
        $M_{\rm max}\,[M_\odot]$ & $86.3$ & $\sigma_{\rm g,high}\,[M_\odot]$ & $6.3$ \\
        $\lambda_{\rm g}$ & $0.4$ & $\gamma$ & $3.3$ \\
        $\lambda_{\rm g,low}$ & $0.8$ & $\kappa$ & $2.9$ \\
        $\delta_m\,[M_\odot]$ & $4.8$ & $z_p$ & $2.5$ \\
    \end{tabular}
    \end{ruledtabular}
\end{table}

Throughout the calculation, luminosity distances and comoving volume elements are evaluated in a flat $\Lambda$CDM cosmology with $H_0=67.66\,{\rm km}\,{\rm s}^{-1}\,{\rm Mpc}^{-1}$ and $\Omega_{m0}=0.30966$, consistent with Planck 2018~\cite{Planck:2018vyg}. We denote the luminosity distance by $D_L(z)$ and the full-sky differential comoving volume by ${\rm d}V_c/{\rm d}z$.

\subsection{Space-based selection}

A SMBHB may spend years to centuries in the milli-Hertz band before coalescence. Its detectability therefore depends not only on its masses and distance but also on its position along the inspiral at the start of the observation. Let $\tau_c$ be the observer-frame time to coalescence at the beginning of a mission of duration $T_{\rm obs}$. At leading order in the inspiral, the observed GW frequency is~\cite{Gerosa:2019dbe}
\begin{equation}
  f(\tau_c) = \frac{5^{3/8}}{8\pi}
  \left[{\cal M}_c(1+z)\right]^{-5/8}
  \tau_c^{-3/8},
  \label{eq:frequency_evolution}
\end{equation}
where ${\cal M}_c=(m_1m_2)^{3/5}/(m_1+m_2)^{1/5}$ is the source-frame chirp mass and $(1+z){\cal M}_c$ is its detector-frame counterpart. All observation and waiting times below are measured in the observer frame.

For the early inspiral, we use the leading-order frequency-domain amplitude, corresponding to the inspiral limit of the nonspinning PhenomA model~\cite{Ajith:2007kx},
\begin{equation}
    |\tilde{h}(f)| = \sqrt{\frac{5}{24}}\,\pi^{-2/3}
    \frac{\left[{\cal M}_c(1+z)\right]^{5/6}}{D_L}f^{-7/6}.
    \label{eq:inspiral_amplitude}
\end{equation}
The amplitude in Eq.~\eqref{eq:inspiral_amplitude} does not include inclination or detector-response factors. We adopt the effective LISA noise power spectral density (PSD), $S_n(f)$, of Ref.~\cite{Robson:2018ifk}, which incorporates the sky- and polarization-averaged response. Averaging additionally over binary inclination gives the accumulated squared signal-to-noise ratio (SNR),
\begin{equation}
  \rho^2(\tau_c) = \frac{16}{5}
  \int_{f_{\rm start}}^{f_{\rm end}}
  \frac{|\tilde{h}(f)|^2}{S_n(f)}\,{\rm d}f.
  \label{eq:SNR_tc}
\end{equation}
The factor $16/5$ is the usual matched-filter factor of $4$ multiplied by the inclination average $4/5$~\cite{Robson:2018ifk}. For an adopted detector band $[f_{\rm low},f_{\rm high}]$, the integration limits are
\begin{equation}
    \begin{aligned}
        f_{\rm start} &= \max[f(\tau_c),f_{\rm low}],\\
        f_{\rm end} &=
        \begin{cases}
            \min[f(\tau_c-T_{\rm obs}),f_{\rm high}], & \tau_c>T_{\rm obs},\\
            f_{\rm high}, & \tau_c\leq T_{\rm obs}.
        \end{cases}
    \end{aligned}
    \label{eq:integration_limits}
\end{equation}
The second branch accounts for sources that coalesce during the observation. We set $\rho=0$ whenever $f_{\rm start}\geq f_{\rm end}$. This prescription includes only the part of the inspiral that lies within both the mission duration and the detector band. Thresholding this angle-averaged SNR provides an approximate space-based selection criterion, rather than a full orientation-dependent detection probability.

\begin{figure}
    \centering
    \includegraphics[width=\linewidth]{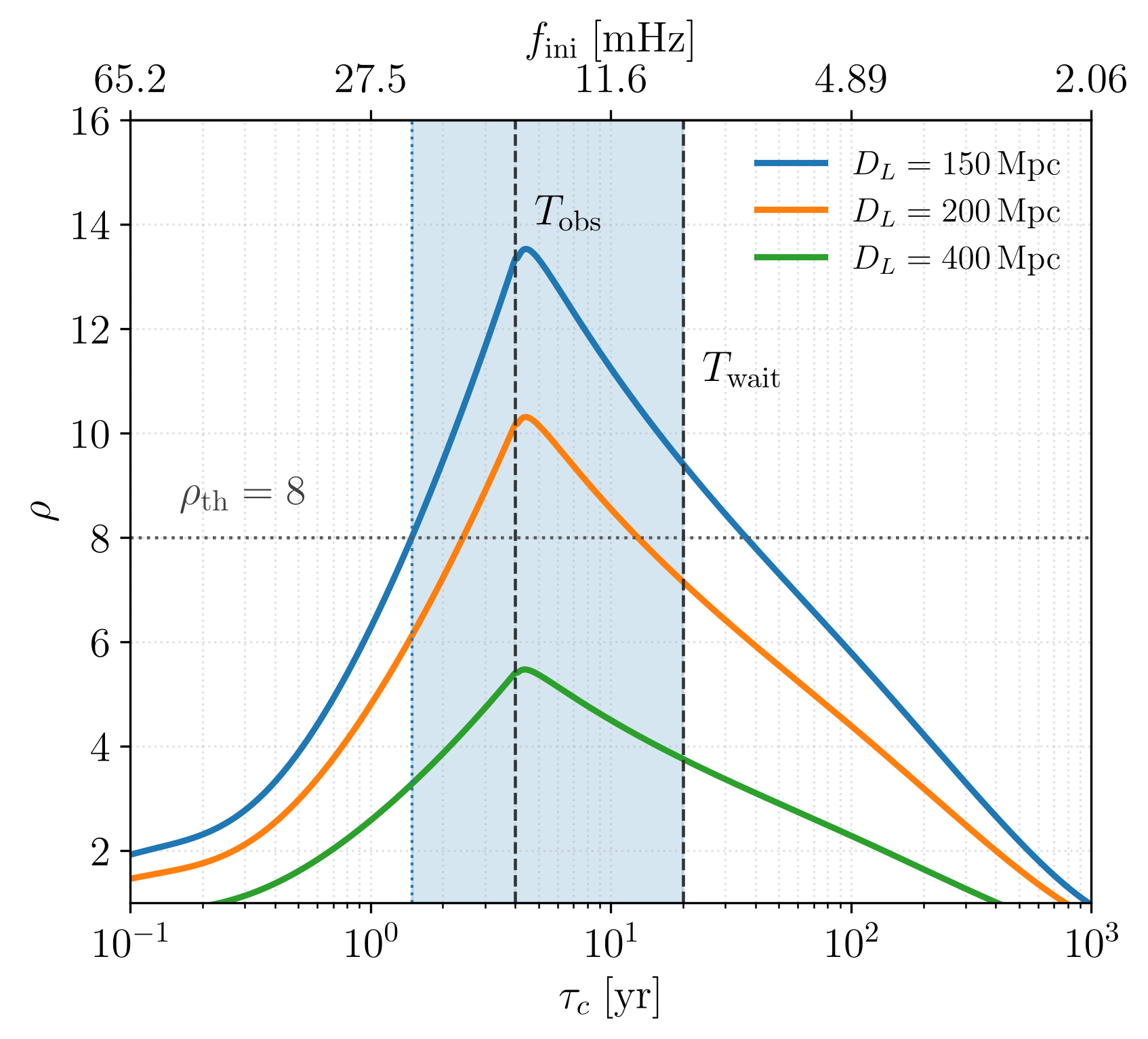}
    \caption{Angle-averaged LISA SNR as a function of the time to coalescence at the start of a $T_{\rm obs}=4\,{\rm yr}$ observation, for binaries with ${\cal M}_c=30\,M_\odot$, $q=1$, and the indicated luminosity distances. The upper axis indicates the initial GW frequency. The horizontal dotted line marks $\rho_{\rm th}=8$, and the vertical dashed lines mark $T_{\rm obs}$ and $T_{\rm wait}=20\,{\rm yr}$. The shaded interval is the space-based detection window of the $D_L=150\,{\rm Mpc}$ source restricted to $\tau_c\leq T_{\rm wait}$; ground-based detectability is accounted for separately.}
    \label{fig:LISA_SNR}
\end{figure}

Fig.~\ref{fig:LISA_SNR} illustrates the dependence of $\rho$ on $\tau_c$ for binaries with ${\cal M}_c=30\,M_\odot$ and $q=1$ at several luminosity distances. At large $\tau_c$, the source radiates at lower frequencies and evolves slowly during the observation; at small $\tau_c$, little time remains in the sensitive part of the LISA band. Consequently, $\rho(\tau_c)$ peaks at an intermediate time to merger. For a detectable system with two threshold crossings, we order the roots of $\rho(\tau_c)=\rho_{\rm th}$ as $\tau_{\rm th1}<\tau_{\rm th2}$ and define the detection-window duration as
\begin{equation}
    \Delta\tau_{\rm space}(m_1,m_2,z)=\tau_{\rm th2}-\tau_{\rm th1}.
    \label{eq:space_window}
\end{equation}
If the maximum SNR does not exceed $\rho_{\rm th}$, the detection window is zero.

We assume that the merger flow is stationary over the time intervals relevant to the observation. At fixed masses and redshift, the number of binaries per unit observer-frame time to merger is then proportional to ${\cal R}(z)/(1+z)$. Integrating this steady merger flow over the detection window gives~\cite{Gerosa:2019dbe}
\begin{align}
    N_{\rm space} ={}& \iiint {\rm d}m_1\,{\rm d}m_2\,{\rm d}z\,
    p(m_1,m_2)\frac{{\cal R}(z)}{1+z}\frac{{\rm d}V_c}{{\rm d}z} \nonumber\\
    &\times\Delta\tau_{\rm space}(m_1,m_2,z).
    \label{eq:N_space}
\end{align}
The factor $(1+z)^{-1}$ converts the source-frame merger rate to an observer-frame rate. The dependence on $T_{\rm obs}$ is already contained in $\Delta\tau_{\rm space}$, so no additional factor of mission duration is required in Eq.~\eqref{eq:N_space}.

\subsection{Multiband selection}

A multiband source must satisfy the space-based selection criterion, coalesce within a specified waiting time, and be detectable by a ground-based instrument. We use a single LIGO detector at design sensitivity as a representative ground-based detector. For this estimate, we compute the optimal SNR using the nonspinning PhenomA inspiral-merger-ringdown waveform~\cite{Ajith:2007kx},
\begin{equation}
    \rho_{{\rm g,opt}}^2 = 4\int_{f_{{\rm g,low}}}^{f_{{\rm g,high}}}
    \frac{|\tilde{h}_{\rm IMR}(f)|^2}{S_{n,\rm g}(f)}\,{\rm d}f,
    \label{eq:ground_snr}
\end{equation}
where $S_{n,\rm g}(f)$ is the ground-based noise PSD, $[f_{{\rm g,low}},f_{{\rm g,high}}]$ is the analysis band, and $\tilde{h}_{\rm IMR}$ is the strain for an optimally located and oriented source. Unlike Eq.~\eqref{eq:SNR_tc}, this optimal SNR contains no angular averaging. PhenomA is used here for approximate SNR forecasts, not for precise parameter estimation.

For a right-angle interferometer, the antenna responses to the two GW polarizations are
\begin{align}
    F_+ ={}& \frac{1}{2}(1+\cos^2\theta)\cos 2\phi\cos 2\psi \nonumber\\
    &-\cos\theta\sin 2\phi\sin 2\psi,\\
    F_\times ={}& \frac{1}{2}(1+\cos^2\theta)\cos 2\phi\sin 2\psi \nonumber\\
    &+\cos\theta\sin 2\phi\cos 2\psi,
\end{align}
where $(\theta,\phi)$ specify the source position in the detector frame and $\psi$ is the polarization angle. For binary inclination $\iota$, the observed SNR is $\rho_{\rm g}=\omega\rho_{{\rm g,opt}}$, with
\begin{equation}
    \omega=\left[\frac{(1+\cos^2\iota)^2}{4}F_+^2
    +\cos^2\iota\,F_\times^2\right]^{1/2},
\end{equation}
and $0\leq\omega\leq1$. Isotropic sky positions and orbital orientations correspond to uniform distributions in $\cos\theta$ and $\cos\iota$ on $[-1,1]$, in $\phi$ on $[0,2\pi)$, and in $\psi$ on $[0,\pi)$. These distributions determine the normalized projection-factor density $p(\omega)$. The probability of exceeding a ground-based threshold $\rho_{{\rm g,th}}$ is then~\cite{Gerosa:2019dbe}
\begin{equation}
    p_{\rm det}(m_1,m_2,z)=
    \int_{\rho_{{\rm g,th}}/\rho_{{\rm g,opt}}(m_1,m_2,z)}^1
    p(\omega)\,{\rm d}\omega,
    \label{eq:ground_selection}
\end{equation}
when $\rho_{{\rm g,opt}}>\rho_{{\rm g,th}}$, and zero otherwise.

We measure the maximum waiting time, $T_{\rm wait}$, from the start of the space-based observation and require $\tau_c\leq T_{\rm wait}$. Following Ref.~\cite{Gerosa:2019dbe}, we take $T_{\rm wait}=5T_{\rm obs}$, corresponding to $20\,{\rm yr}$ for a four-year mission. The overlap of the space-based detection window with this waiting-time interval is
\begin{equation}
    \Delta\tau_{\rm multi}
    =\max\!\left\{0,\min[\tau_{\rm th2},T_{\rm wait}]-\tau_{\rm th1}\right\},
    \label{eq:multi_window}
\end{equation}
with $\Delta\tau_{\rm multi}=0$ for systems that never pass the space-based threshold. An example is shown by the shaded region in Fig.~\ref{fig:LISA_SNR}.

Within our angle-averaged treatment of space-based selection, we approximate the joint selection by multiplying this time window by $p_{\rm det}$. The expected multiband count is therefore
\begin{align}
    N_{\rm multi} ={}& \iiint {\rm d}m_1\,{\rm d}m_2\,{\rm d}z\,
    p(m_1,m_2)\frac{{\cal R}(z)}{1+z}\frac{{\rm d}V_c}{{\rm d}z} \nonumber\\
    &\times p_{\rm det}(m_1,m_2,z)\,
    \Delta\tau_{\rm multi}(m_1,m_2,z).
    \label{eq:N_multi}
\end{align}
We assume that the ground-based detector is available at the merger time, corresponding to a $100\%$ duty cycle. This prescription does not model correlations between the orientation-dependent selection functions of the space- and ground-based instruments.

\subsection{Detection forecasts}\label{sec:detection_forecasts}

We evaluate Eqs.~\eqref{eq:N_space} and~\eqref{eq:N_multi} by Monte Carlo integration. For a four-year space-based observation with $\rho_{\rm th}=8$, and adopting the local SMBHB merger-rate density $R_0=19^{+7}_{-5}\,{\rm Gpc}^{-3}\,{\rm yr}^{-1}$ inferred from GWTC-4~\cite{LIGOScientific:2026vua}, we obtain
\begin{equation}
    N_{\rm space}=4.64^{+1.71}_{-1.22}.
\end{equation}
Here and below, the quoted uncertainties propagate only the uncertainty in $R_0$, with all other population hyperparameters held fixed. They therefore do not include the full population-model uncertainty or Poisson fluctuations in the realized number of detections.

\begin{figure}
    \centering
    \includegraphics[width=\linewidth]{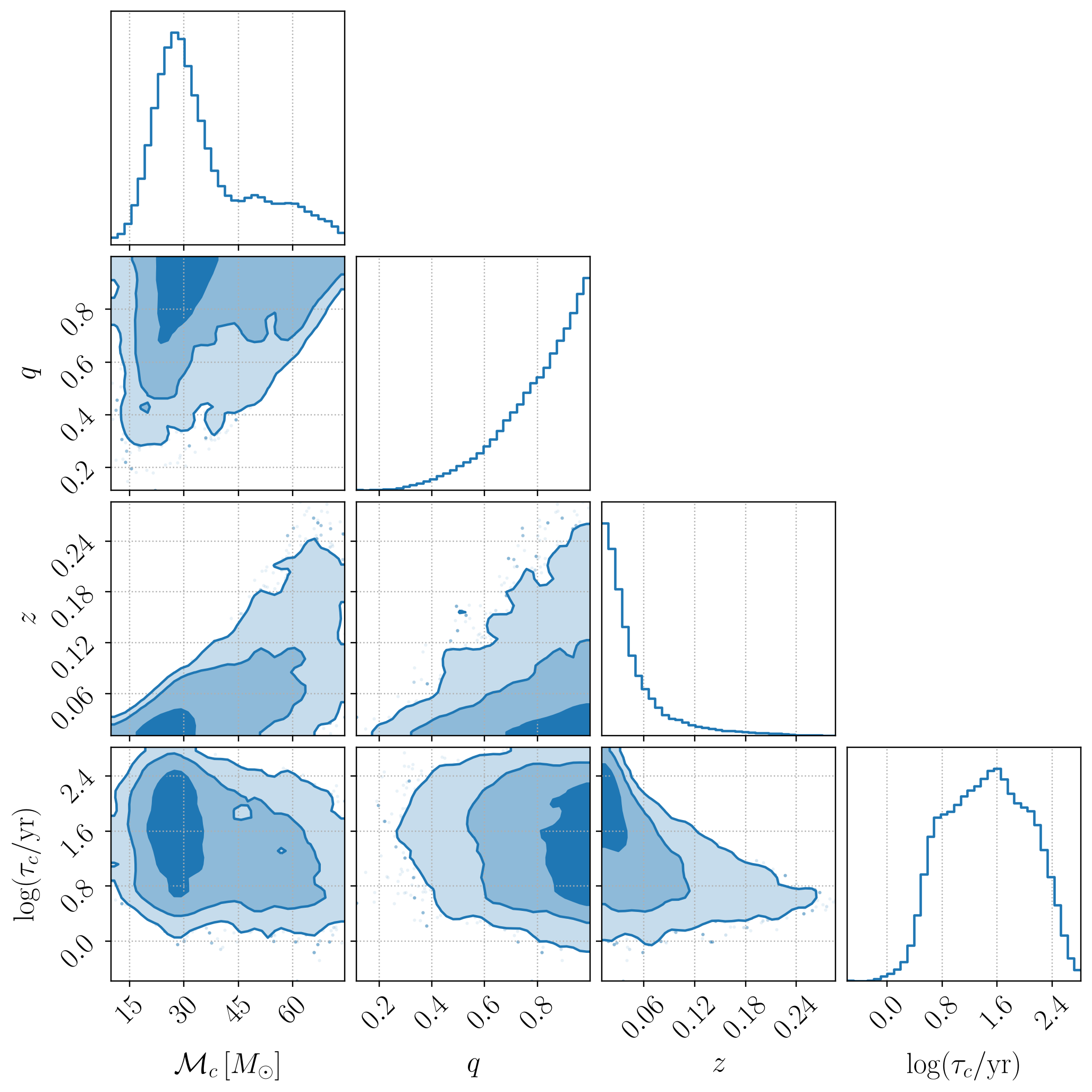}
    \caption{Forecast distributions of source parameters for SMBHBs detected by LISA with $T_{\rm obs}=4\,{\rm yr}$ and $\rho_{\rm th}=8$, using the population model in Sec.~\ref{sec:source_population}. Masses are defined in the source frame, and $\tau_c$ is the observer-frame time to coalescence at the start of the observation.}
    \label{fig:lisa_param_dist}
\end{figure}

The corresponding distributions of detected-source parameters are shown in Fig.~\ref{fig:lisa_param_dist}. The chirp-mass distribution peaks near ${\cal M}_c\simeq30\,M_\odot$ and has a tail toward higher masses, while the mass-ratio distribution favors nearly equal-mass systems. The forecast sample is dominated by nearby binaries, with $z\lesssim0.3$, and typical times to coalescence of $10$-$100\,{\rm yr}$. Thus, space-based detection does not in general imply a merger during the mission or within the adopted multiband waiting time. These qualitative trends are consistent with previous studies of SMBHBs in LISA~\cite{Gerosa:2019dbe,Buscicchio:2024asl}.

Requiring a merger within $T_{\rm wait}=20\,{\rm yr}$ and imposing $\rho_{{\rm g,th}}=8$ for a LIGO detector at design sensitivity gives
\begin{equation}
    N_{\rm multi}=1.42^{+0.52}_{-0.37},
\end{equation}
approximately $30\%$ of the space-based sample at $\rho_{\rm th}=8$. Ground-based detections can also guide targeted searches of archival LISA data, reducing the search volume and allowing weaker space-based signals to be recovered~\cite{Wong:2018uwb}. To illustrate this possibility, we consider a space-based threshold of $\rho_{\rm th}=5$ while retaining $\rho_{{\rm g,th}}=8$ and the same waiting time. The expected multiband count then becomes
\begin{equation}
    N_{\rm multi}=6.65^{+2.45}_{-1.75}.
\end{equation}
This is an illustrative threshold-based forecast rather than a calibrated recovery efficiency for a specific search pipeline. Because it uses a lower LISA threshold, this scenario draws on a larger space-based sample than the $\rho_{\rm th}=8$ forecast. Ground-informed searches may therefore enlarge the sample available for multiband tests of cloud-induced orbital dynamics.

\section{Observational signatures of cloud level transitions}
\label{sec:signatures}

In this section, we discuss the effects of superradiant cloud level transition on SMBHB dynamics and their GW radiation, and the potential signatures in milli-Hertz GW observation. We shall start with figuring out the parameter space of the relevant transitions, and then compute phase drift in GWs and delay in merger time caused by superradiant cloud.

\subsection{Parameter space}\label{sec:para}

Let's consider an isolate black hole of mass $M$ and with initial dimensionless spin $a_i$. The time for a superradiant mode $\ket{n\ell m}$ to saturate can be estimated as
\be
\tau^{\rm sat}_{n \ell m} = \frac{\ln \left({N^{\rm sat}_{n\ell m}}/{N^{\rm seed}_{n\ell m}}\right)}{2\Gamma_{n\ell m}} \, ,
\ee
where $N^{\rm seed}_{n\ell m}$ and $N^{\rm sat}_{n\ell m}$ are the initial and saturation occupation number of the mode, $\Gamma_{n\ell m} = {\rm Im} \omega_{n\ell m}$ with $\omega_{n\ell m}$ being the eigenfrequency of $\ket{n\ell m}$. For $\ket{211}$, we have
\be
N^{\rm sat}_{211} \simeq G M^2 (a_i - 4\alpha) \,,
\ee
and the corresponding growth time is
\be
\tau^{\rm sat}_{211} \approx 10^7 {\rm yrs} \times \left(\frac{M}{35 M_\odot}\right)  \left(\frac{0.024}{\alpha}\right)^9 \,,
\ee
where we take $a_i = 0.99$ and $N_{211}^{\rm seed} = 1$. After superradiant growth, $\ket{211}$ depletes by radiating GWs, which takes
\be
\tau^{\rm GW}_{211}  \approx 10^7 {\rm yrs} \times \left(\frac{M}{35 M_\odot}\right)  \left(\frac{0.08}{\alpha}\right)^{15} \,.
\ee
On the other hand, $\ket{322}$ also grow superradiantly but typically grows much slowly comparing to $\ket{211}$. Hence, the saturation occupation number of $\ket{322}$ can be estimated as
\be
N^{\rm sat}_{322} \simeq \frac{G M^2}{2} \left(a^{\rm sat}_{211} -a^{\rm sat}_{322} \right) \approx \alpha G M^2 \,,
\ee
and the growth time is 
\be
\tau^{\rm sat}_{322} \approx 10^7 {\rm yrs} \times \left(\frac{M}{35 M_\odot}\right)  \left(\frac{0.14}{\alpha}\right)^{14} \,.
\ee
After saturation, $\ket{322}$ depletes by GW radiations on a time scale of
\be
\tau^{\rm GW}_{322}  \approx 10^7 {\rm yrs} \times  \left(\frac{M}{35 M_\odot}\right)  \left(\frac{0.28}{\alpha}\right)^{20} \,.
\ee

If the black hole is a binary system, the cloud also evolve under the tidal potential of the companion object, and may undergo resonant transition at certain orbits. It is convenient to use $f = \Omega/\pi$, which is the frequency of GWs radiated by a circular orbit of orbital frequency $\Omega$, to denote the orbit. For simplicity, we assume the BH spin is align with the orbital angular momentum. In this case, the $\ket{211}$-mode may undergo hyperfine, fine and Bohr resonant transitions at
\be
\ba
&f_{211}^{\rm HF}  \approx 0.01 {\rm Hz} \times   \left(\frac{35 M_\odot}{M}\right) \left(\frac{\alpha}{0.21}\right)^7  \\
&f_{211}^{\rm F}  \approx 0.01 {\rm Hz} \times   \left(\frac{35 M_\odot}{M}\right) \left(\frac{\alpha}{0.10}\right)^5  \\
&f_{211}^{\rm B}  \approx 0.01 {\rm Hz} \times   \left(\frac{35 M_\odot}{M}\right) \left(\frac{\alpha}{0.024}\right)^3  \\
\ea
\ee
and transits to $\ket{21-1}$, $\ket{200}$ and $\ket{100}$ modes respectively. In the calculation, we take $a^{\rm sat}_{211} = \frac{4\alpha}{1+4\alpha^2} \approx 4\alpha$. Similarly, the $\ket{322}$-mode may undergo hyperfine, fine and Bohr resonant transitions at
\be
\ba
&f_{322}^{\rm HF}  \approx 0.01 {\rm Hz} \times   \left(\frac{35 M_\odot}{M}\right) \left(\frac{\alpha}{0.34}\right)^7  \\
&f_{322}^{\rm F}  \approx 0.01 {\rm Hz} \times   \left(\frac{35 M_\odot}{M}\right) \left(\frac{\alpha}{0.22}\right)^5  \\
&f_{322}^{\rm B}  \approx 0.01 {\rm Hz} \times   \left(\frac{35 M_\odot}{M}\right) \left(\frac{\alpha}{0.043}\right)^3  \\
\ea
\ee
and transits to $\ket{320}$, $\ket{31-1}$ and $\ket{211}$ modes respectively.\footnote{The $\ket{322}$ mode may also transit to $\ket{32-2}$, but with less efficiency, and hence is not considered here.} In the calculation, we take $a^{\rm sat}_{322} = \frac{2\alpha}{1+\alpha^2} \approx 2\alpha$.

It is also useful to list the lifetime of the resonant orbits (assuming the orbits decay by radiating GWs, without superradiant cloud backreaction on the orbital dynamics). For $\ket{211}$ and $\ket{322}$, the corresponding orbital lifetime can be estimated as 
\be
\ba
& \tau_{211}^{\rm HF}  \approx 10^7 {\rm yrs} \times \left(\frac{M}{35 M_\odot}\right) \left(\frac{\alpha}{0.10}\right)^{56/3} \\
& \tau_{322}^{\rm HF}  \approx 10^7 {\rm yrs} \times \left(\frac{M}{35 M_\odot}\right) \left(\frac{\alpha}{0.17}\right)^{56/3} 
\ea
\ee

The age of a stellar mass black hole is typically longer than $10^7$ years. For simplicity, we shall assume the age of the black hole is $10^7$ years. Under this assumption, we consider the parameter space with
\be\label{con1}
\tau^{\rm sat}_{n\ell m} < 10^7 {\rm yrs} < \tau^{\rm GW}_{n\ell m}\,,
\ee
such that the superradiant mode sufficiently grows and does not deplete by GW radiation before level transitions.\footnote{Strictly speaking, the cloud does not have to reach saturation before level transitions. As will be shown later, the detectability of the cloud mass can reach $10^{-6}$, which is far below the saturation mass. However, given the sensitive dependence on $\alpha$, removing this constraint would not impact the parameter space very much.} \footnote{If the resonant orbits are lower than the initial orbit, the lifetime of the resonant orbits is typically much shorter than that of the initial orbits.} The age of the black hole which also defines the lifetime of the initial orbit as we are interested in binaries that mergers today. In order to have a certain resonant transition, we require the lifetime of the corresponding resonant orbit is shorter than the age of the black hole and the saturation time 
\be
\tau^{\rm HF, F, B}_{n\ell m} < 10^7 {\rm yrs} \,.
\ee

\begin{figure}[t]
    \centering
    \includegraphics[width=\columnwidth]
    {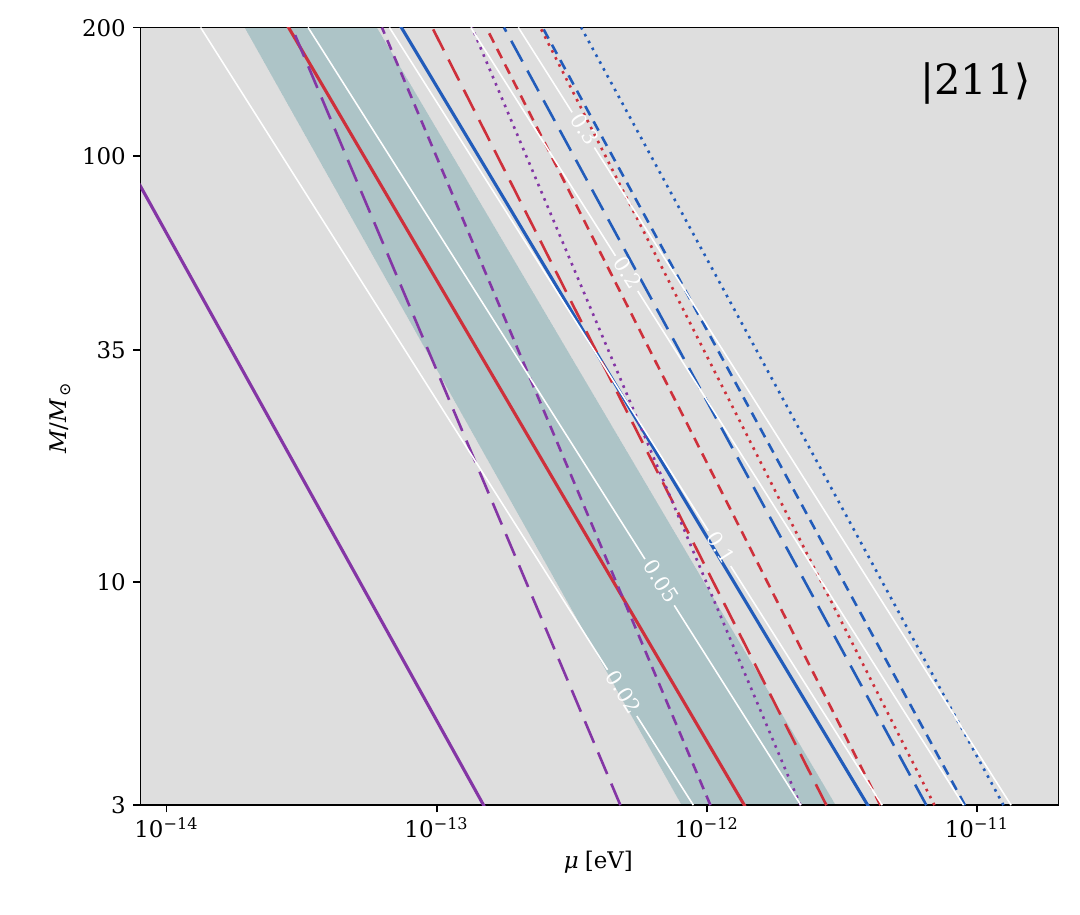}
    \vspace{0.5em}
    \includegraphics[width=\columnwidth]
    {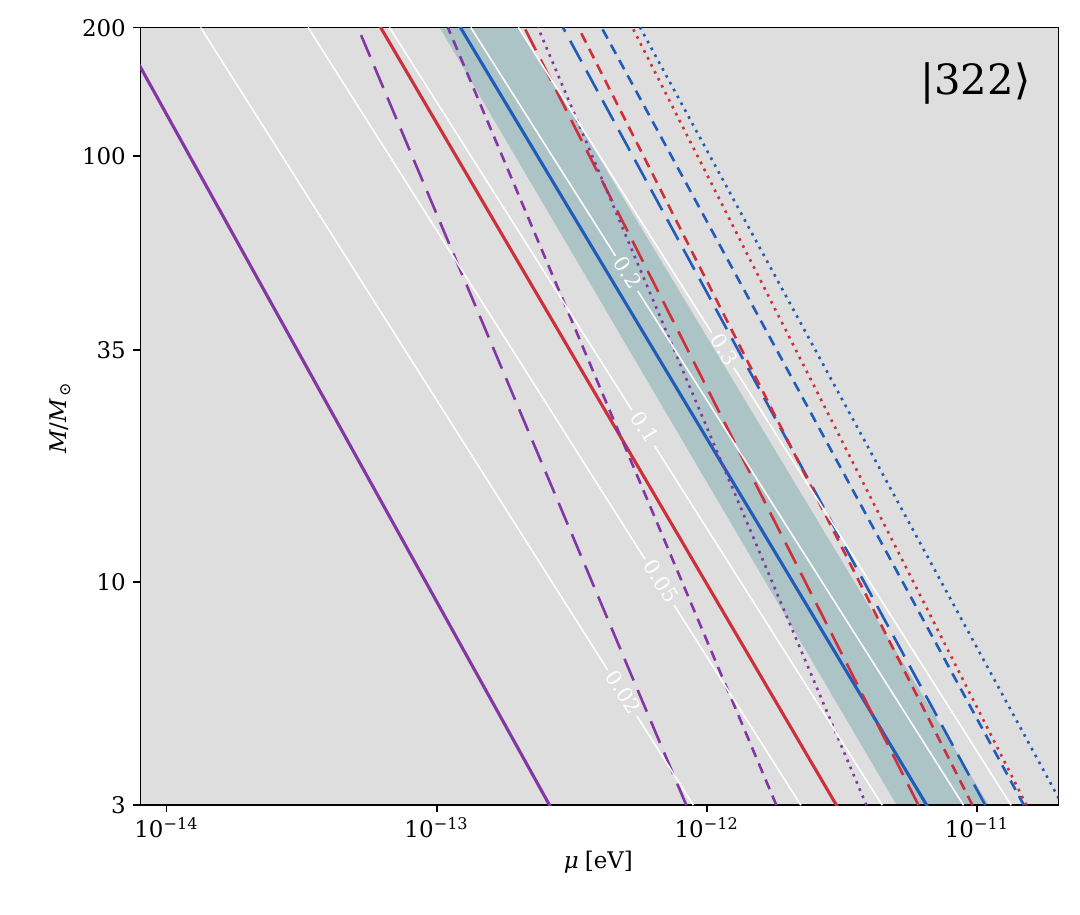}
    \caption{Parameter space of resonant level transitions. The upper and lower plots show the parameter space of $\ket{211}$ and $\ket{322}$ respectively. The green region show the parameter space satisfying $\tau^{\rm sat}_{n\ell m} < 10^7 {\rm yrs} < \tau^{\rm GW}_{n\ell m}$. The colored solid lines are the contours of $\tau^{\rm HF, F, B}_{n\ell m} = 10^7 {\rm yrs}$, with blue, red and purple for the hyperfine, fine and Bohr resonance respectively. The long dashed lines, the dashed lines and the dotted lines show the frequency of GWs emitted at the resonant orbits, and correspond to $0.01$, $0.1$ and $1$ Hz respectively. The plots also show the contours of $\alpha$ in thin white lines.}
    \label{fig:para}
\end{figure}
In Fig.~\ref{fig:para}, we plot the corresponding parameter space. It shows that for $\ket{211}$, subjecting to condition.~\eqref{con1}, hyperfine resonance typically happens at orbits much higher than the initial orbit (here a higher orbit means the orbit has larger orbital radius), and hence is not relevant.  While fine resonance could happen at orbits lower than the initial orbit, namely after the black hole formation, it typically happens before the binary enters the LISA-band, and hence is also not relevant for the observation. Bohr resonance can happen when binaries are within the LISA band, however, the cloud may not grow sufficiently when the binary enter the LISA band for black hole with mass lower than $35 M_\odot$. Nevertheless, the resonance can alter the binary merger time, and might be inferred with multiband observations, even if it cannot be observed directly by LISA. This transition can also be potential targets for deci-Hertz GW detections. For $\ket{322}$, hyperfine resonance typically happens before the binary enters the LISA band, and cannot be detected. The fine resonance can happen in the LISA band and is interesting for milli-Hz GW observations. The Bohr resonance also happens after the binary passing the LISA band, and could be inferred with multiband observations and deci-Hertz GW observations.

\subsection{Orbital dynamics}

We shall consider a transition from state $\ket{a}$ to $\ket{b}$. Here $a$ and $b$ are the shorthand for the quantum numbers of the states. The destination state can be a bounded state or an unbounded state. In the later case, the quantum numbers are $k_b \ell_b m_b$ with $k_b$ being the wavenumber of the unbounded state. The tidal coupling strength between two states $\ket{a}$ and $\ket{b}$ is defined as
\be
\eta_{ba} = \bra{b} V \ket{a} {e^{i(m_a-m_b)\Omega t}} \,,
\ee
where $V$ is the gravitational potential of the companion object. As discussed in Sec.~\ref{sec:para}, we shall focus on the fine and Bohr transitions. While these transitions are between bound states, cloud ionization, i.e., transition from bound states to unbounded states, might also take places during Bohr transitions, and affect both cloud and orbit evolution. Hence, evolution of the cloud mass $M_c$ and the orbital frequency $\Omega$ can be generally described by
\begin{eqnarray}
\dot{M}_c &=&  \Gamma_{\rm res} M_c + \Gamma_{\rm ion} M_c\,, \label{eq:dMc} \\
\dot{\Omega} &=& \gamma_{\rm GW} + \gamma_{\rm res} + \gamma_{\rm ion} \,, \label{eq:dOmega}
\end{eqnarray}
where 
\be
\gamma_{\rm GW} = \frac{96}{5} \frac{q}{(1+q)^{1/3}} (GM\Omega)^{5/3}\Omega^2 \,, \\
\ee
is the orbital decay rate caused by GW radiation, and $\Gamma_{\rm res}$ and $\gamma_{\rm res}$ ($\Gamma_{\rm ion}$ and $\gamma_{\rm ion}$) denote the backreaction of cloud level transition (ionization) on the cloud mass and orbital dynamics. When the coupling strength is much weaker than the decay rate of the target mode, i.e., $|\eta_{ba}| \ll |\Gamma_b|$, particles transition to $\ket{b}$ can be absorbed by the black hole immediately, and the occupation number in $\ket{b}$ can reaches equilibrium state during the transition. In this case, we have
\be
\ba
&\Gamma_{\rm res} =  \frac{2 \Gamma_b\eta_{ba}^2}
{(m_a-m_b)^2(\Omega-\Omega_{0})^2+\Gamma_b^2} \,,\\
&\Gamma_{\rm ion} = -\sum_{\ell_c,m_c} \frac{\mu |\eta_{ca}|^2}{k_c} \Theta(k_c^2) \, ,\\
&\gamma_{\rm res} =  \frac{3(1+q)^{1/3}}{q\alpha} (m_a-m_b)\Omega_0 (GM\Omega)^{1/3} \frac{M_c}{M} \Gamma_{\rm res} \,, \\
&\gamma_{\rm ion} = \frac{3(1+q)^{1/3}}{q\alpha} (GM\Omega)^{1/3}\Omega \frac{M_{\rm c}}{M} \sum_{\ell_c, m_c} (m_c-m_a) \frac{\mu |\eta_{ca}|^2}{k_c} \Theta(k_c^2) \ \,,
\ea
\ee
where $\Omega_0$ is the resonant orbit frequency, $\Theta$ is the step function. In principle, the summations in $\Gamma_{\rm ion}$ and $\gamma_{\rm ion}$ are over all unbounded states $\ket{c}$ subject to the selection rule. In the later numerical computation, we only include the states with $|m_c-m_a| \le 20$, as states with higher $|m_c-m_a|$ do not contribute significantly.

Combining Eqs.~\eqref{eq:dMc} and~\eqref{eq:dOmega}, we have
\be
M_c(\Omega) \simeq M_c^{i} \exp\left( \int  \frac{\Gamma_{\rm res}+\Gamma_{\rm ion}}{\gamma_{\rm GW}+\gamma_{\rm res}+\gamma_{\rm ion}}{\rm d}\Omega\right) \,,
\ee
where $M_c^{i}$ is the initial mass of the cloud. Eq.~\eqref{eq:dOmega} also indicates that there will be floating orbits at $\Omega_{\rm flt}$, if $\gamma_{\rm GW}(\Omega_{\rm flt}) + \gamma_{\rm ion}(\Omega_{\rm flt}) = -\gamma_{\rm res}(\Omega_{\rm flt})$. 

\begin{figure}[t]
    \centering
    \includegraphics[width=\columnwidth]
    {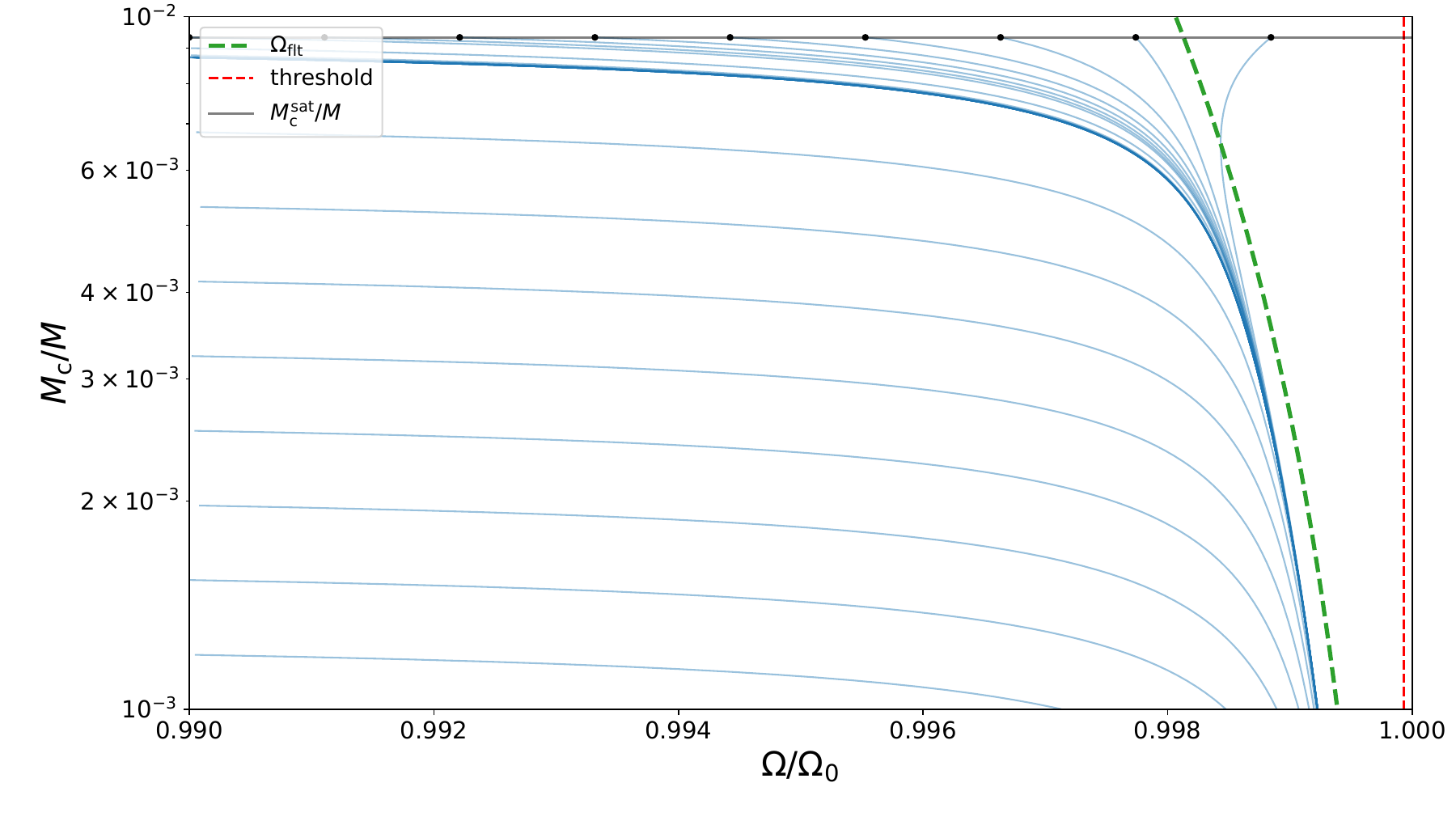}
    \vspace{0.5em}
    \includegraphics[width=\columnwidth]
    {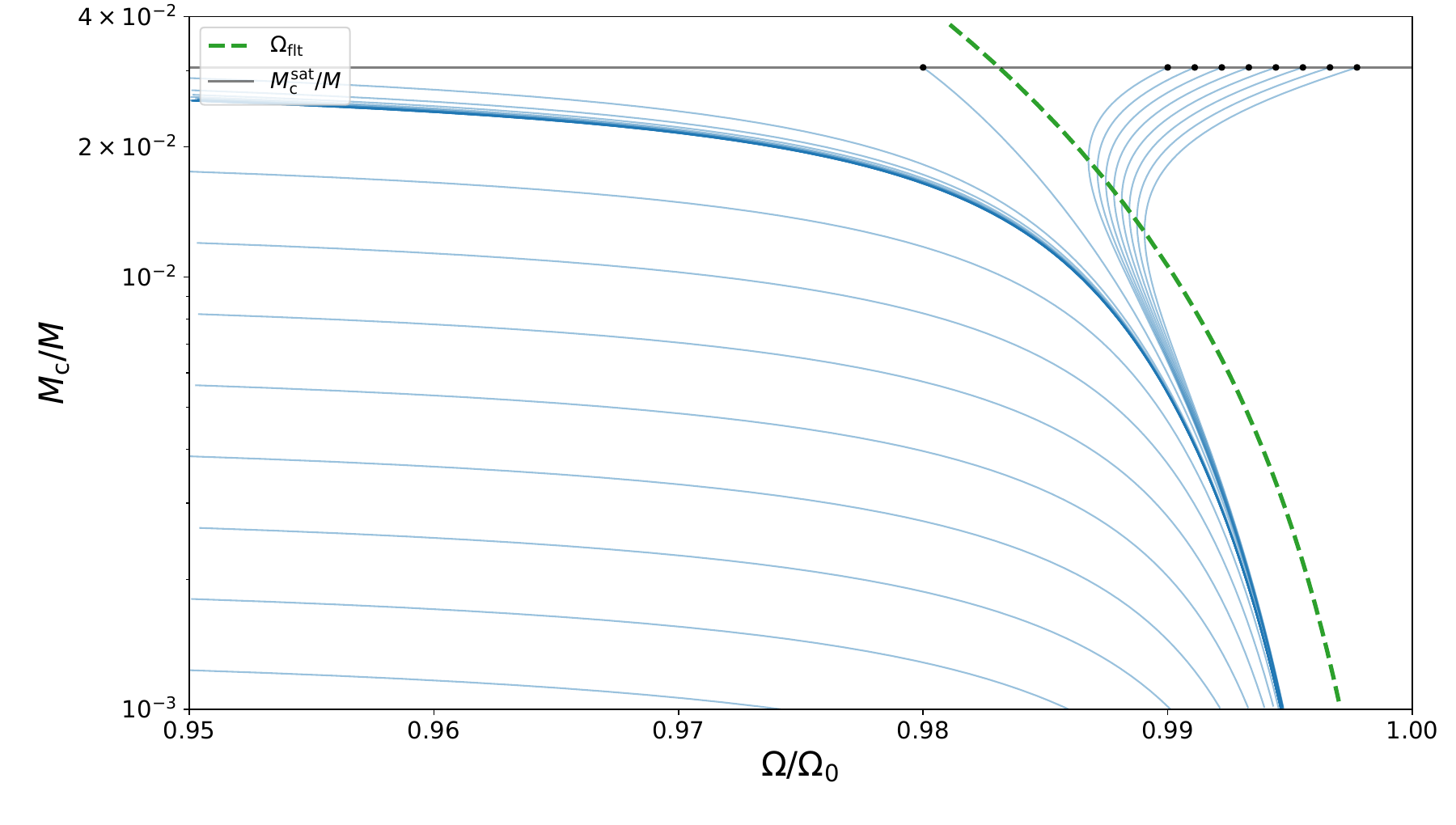}
    \caption{Evolution of orbital frequency and cloud mass near the $\ket{322}$ to $\ket{31-1}$ resonance. We take $q=1$ and $\alpha = 0.1$ and $0.2$ in the upper and lower panels respectively. The horizontal gray lines show the saturation cloud mass, the green dashed lines show the floating orbits. We also mark the condition $|\eta_{322}^{\rm F}| = |0.2 \Gamma_{322}|$ with red dashed lines. As shown in the plots, all solutions considered here satisfy the condition $|\eta_{322}^{\rm F}| < 0.2 | \Gamma_{322}|$.}
    \label{fig:tr322F}
\end{figure}

For demonstration, we first consider the fine transition from $\ket{322}$ to $\ket{31-1}$, in which case the orbital separation is much larger than the Bohr radius of the cloud such that cloud ionization can be neglected, and the coupling strength is given by
\be
\eta_{322}^{\rm F} = \frac{2025}{4} \frac{q}{(1+q)^{4/3}} \frac{(GM\Omega)^{5/3}\Omega}{\alpha^5} \,.\\
\ee
The evolution near $\ket{322}$-to-$\ket{31-1}$ fine resonance can be obtained by solving Eqs.~\eqref{eq:dOmega} and~\eqref{eq:dMc} numerically. Fig.~\ref{fig:tr322F} shows the numerical solutions with different initial conditions. We find that the cloud typically starts to deplete before the orbit reaches the resonant orbit. It also shows that if a cloud form after the orbit passed the floating orbits, the orbit may first outspiral and then continue decay after the cloud mass drops below certain value.

For the Bohr transitions, we have 
\be
\ba
&\eta_{322}^{\rm B} = \frac{q\alpha^3}{GM} \frac{35831808}{1953125x^4}\left[1-e^{-5x/6}Q_7(x)\right], \\
&\eta_{211}^{\rm B} = \frac{q\alpha^3}{GM}\, e^{-3x/2} \left( \frac{2x}{9} +\frac{16}{27} +\frac{64}{81x} +\frac{128}{243x^2} \right),
\ea
\ee
where $x = (1+q)^{1/3}\alpha^2/(GM\Omega)^{2/3}$ and
\be
\ba
Q_7(x)=&1+\frac56 x +\frac{25}{54}x^2+\frac{125}{648}x^3 \\
&+\frac{625}{10368}x^4 +\frac{15625}{1119744}x^5 \\
&+\frac{15625}{6718464}x^6 + \frac{78125}{322486272}x^7.
\ea
\ee
In these cases, cloud ionization might be efficient and has to be taken into account.

\subsection{GW Phase shift}

\begin{figure}[t]
    \centering
    \includegraphics[width=\columnwidth]
    {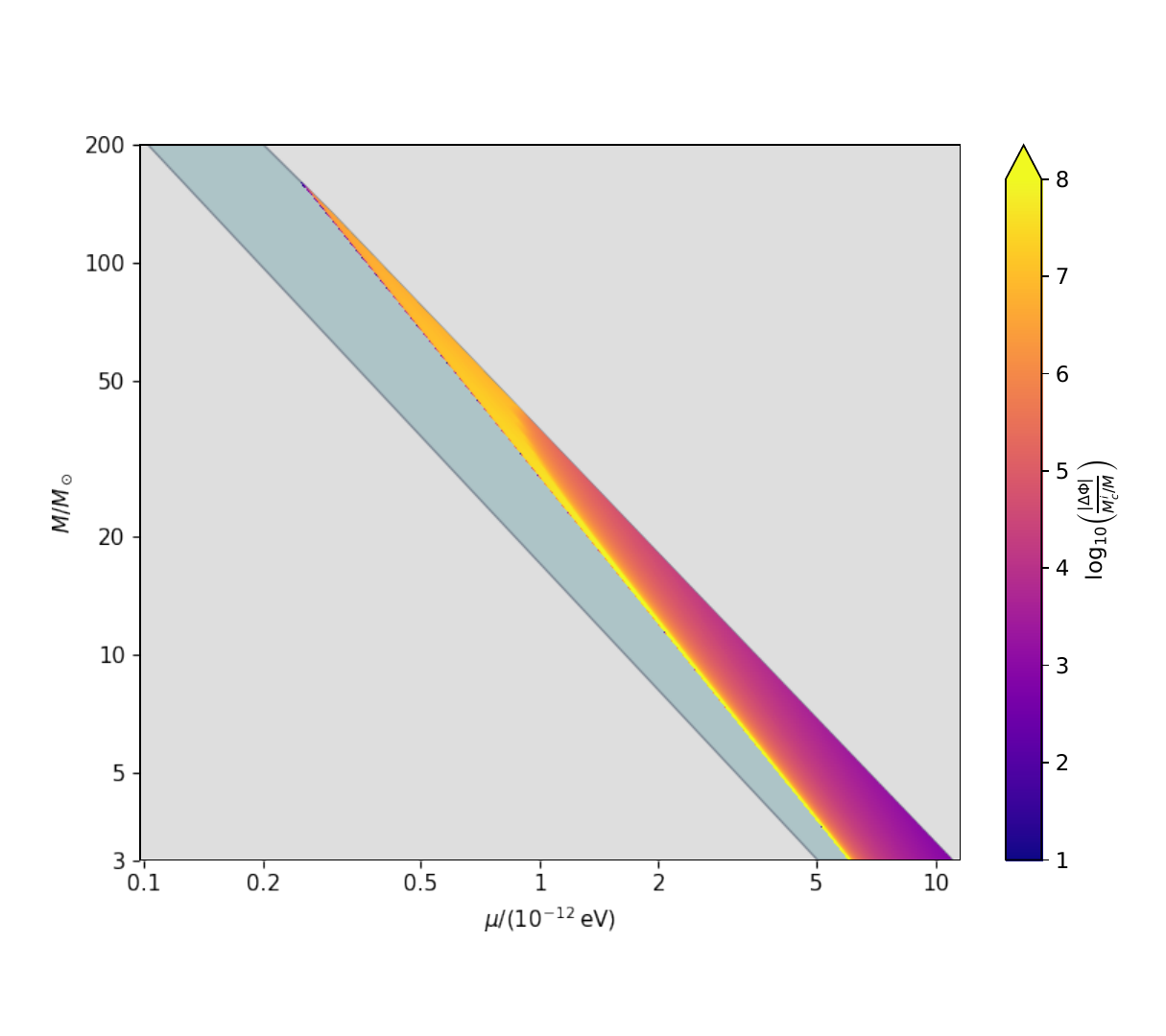}
      \caption{GW phase shift caused by the $\ket{322}$-to-$\ket{31-1}$ transition. The color bar shows the phase shift rescaled by ratio between the initial cloud mass and the black hole mass $M_c^i/M$. We consider the parameter region satisfies $\tau^{\rm sat}_{322} < 10^7 {\rm yrs} < \tau^{\rm GW}_{322}$ such that superradiant cloud can form before level transitions. We take $q=1$ as benchmark, and assume that the binaries are initially detected at $0.01$ Hz and are observed for 10 years. We also assume the cloud mass is small enough such that its effects on the orbital dynamics is perturbative, and hence the phase shift is proportional to the initial cloud mass $M_c^i$.}
    \label{fig:DPhi}
\end{figure}

If the transition happens in the LISA band, it can be detected with the GWs emitted by binaries. Given the orbital evolution, we shall now compute the GW phase shift caused by the superradiant cloud level transition. The insprial GW waveform can be computed using the stationary phase approximation. Specially, waveform in the frequency domain can be written as,
\be
\tilde{h}(f)=\mathcal{A}(f) e^{i\,\Phi(f)},
\ee
where
\be
\Phi = 2\pi f t -\phi-\frac{\pi}{4}\,,
\ee
with
\be
t(f) = t_c - \int \frac{1}{\dot{\Omega}} {\rm d}\Omega\,,
\ee
and
\be
\phi(f) = \phi_c - 2 \int \frac{\Omega}{\dot{\Omega}} {\rm d}\Omega\,.
\ee
Here $t_c$ is the coalescence time, and $\phi_c$ is the phase at $t_c$. When $M_c$ is small such that $\gamma_{\rm res} \ll \gamma_{\rm GW}$, the phase shift induced by level transition can be estimated as
\be\label{eq:DPhi}
\Delta \Phi \simeq - 2\pi f \int \frac{\gamma_{\rm res}+\gamma_{\rm ion}}{\gamma_{\rm GW}^2} {\rm d}\Omega + 2 \int \frac{ \gamma_{\rm res}+\gamma_{\rm ion}} {\gamma_{\rm GW}^2} \Omega {\rm d}\Omega \,,
\ee
with $M_c$ in the above integrals giving by
\be\label{eq:Mcapp}
M_c \simeq M_c^{i} \exp\left( \int  \frac{ \Gamma_{\rm res}+\Gamma_{\rm ion}}{\gamma_{\rm GW}}{\rm d}\Omega\right) \,.
\ee
To the leading order correction, the phase shift is proportional to the initial mass of the cloud $M_c^i$. In Fig.~\ref{fig:DPhi}, we consider the $\ket{322}$-to-$\ket{31-1}$ transition, and calculate the phase shift in GWs by integrating Eq.~\eqref{eq:DPhi} numerically. We assume the GW signals start at $0.01{\rm Hz}$ and are observed for 10 years. The colored region shows the parameter space where the level transition could be detected. If the GW detector can distinguish signals with $|\Delta \phi| \sim {\cal O}(1)$, then one can detected cloud mass with an accuracy of $M^i_c/M < |\Delta \phi|$. In the gray region, the level transition mostly does not affect orbits with $f < 0.01 {\rm Hz}$ and cannot be detected given our assumptions on the targeting binaries. Note that parameter in the gray region can still be probed with lower frequency binaries.

\subsection{Merger time}

\begin{figure}[t]
    \centering
    \includegraphics[width=\columnwidth]
    {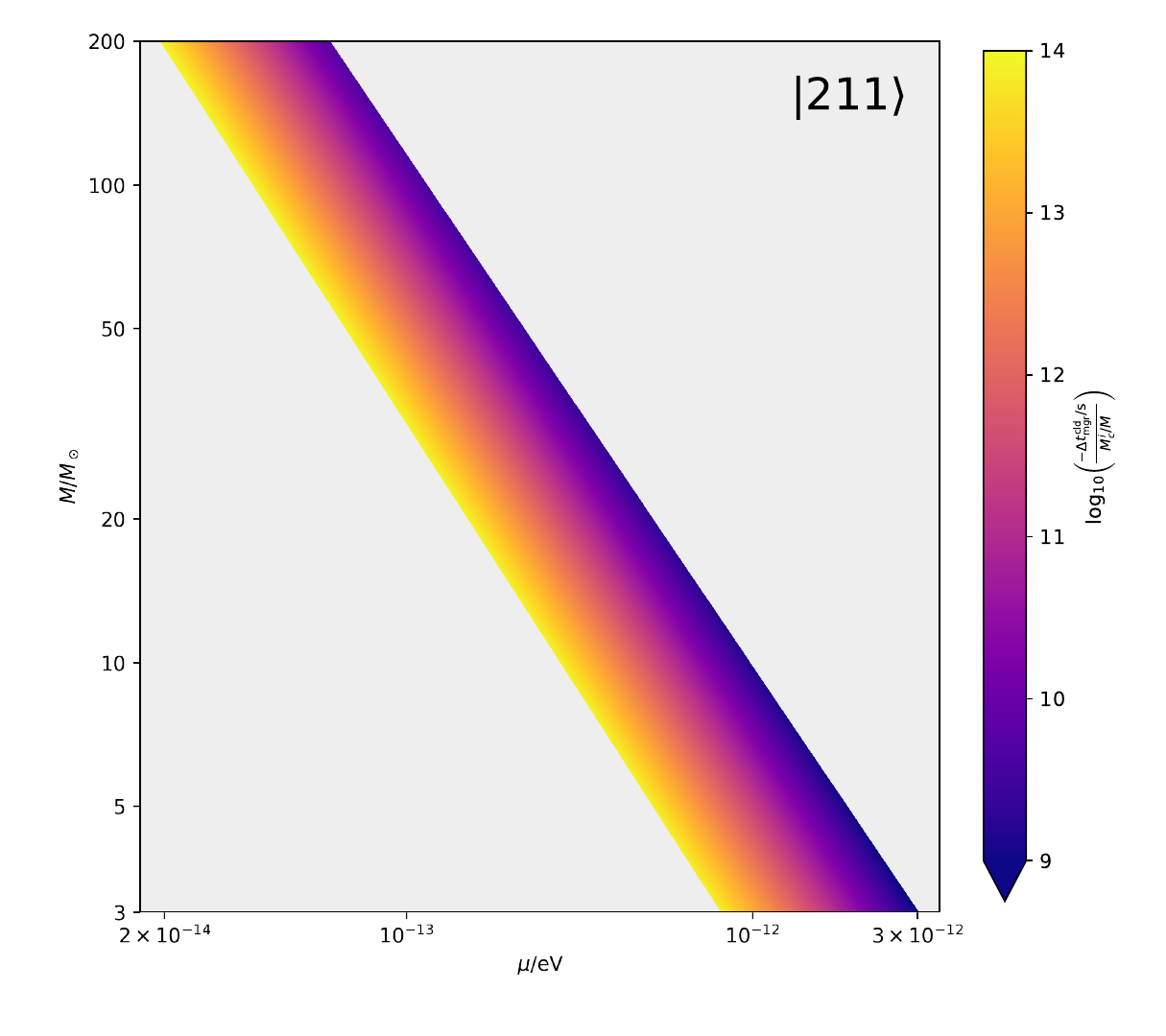}
    \vspace{0.5em}
    \includegraphics[width=\columnwidth]
    {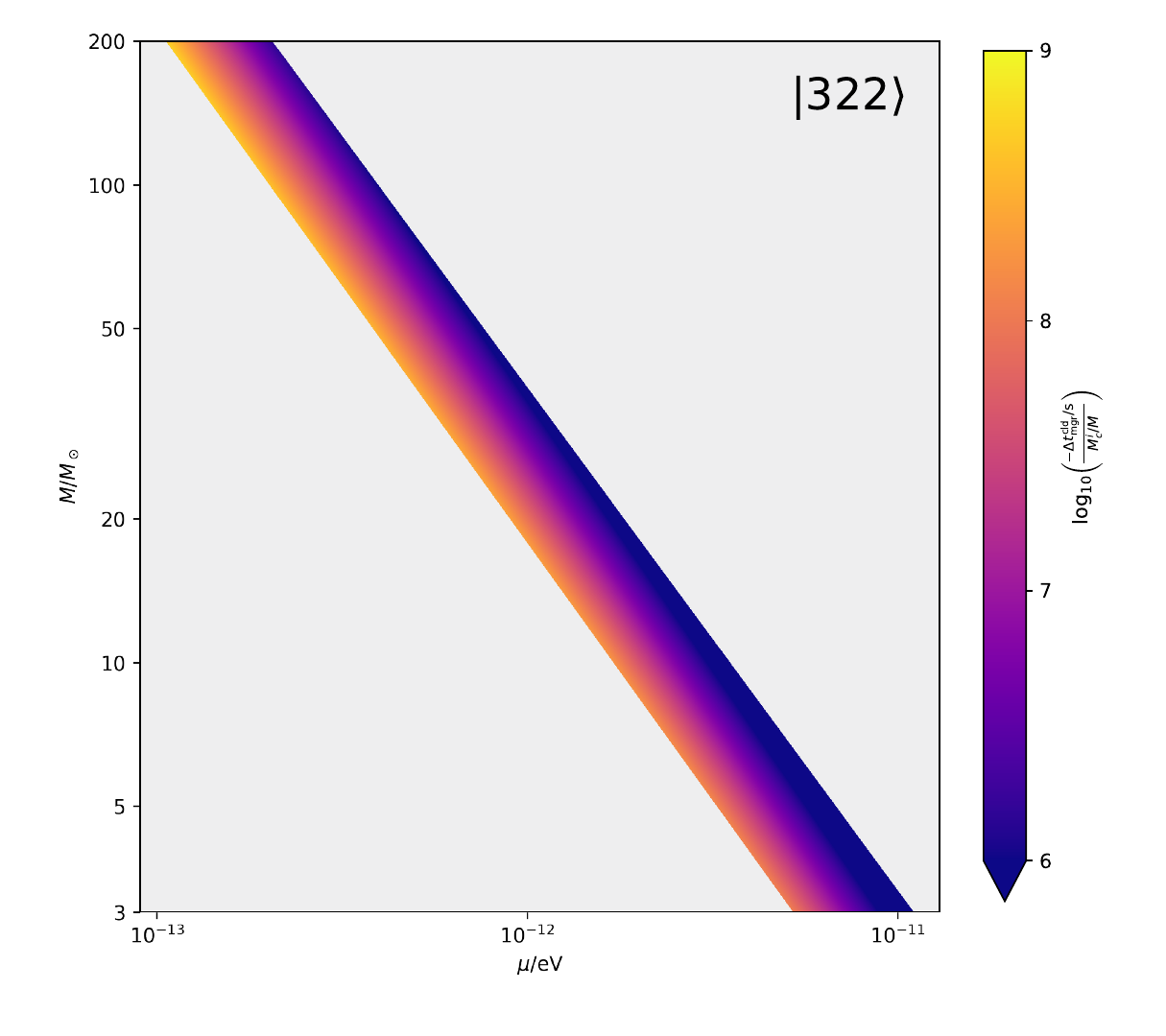}
    \caption{Corrections on the merger time. The upper and lower plots show the corrections on the merger time of $\ket{211}$-to-$\ket{100}$ and $\ket{322}$-to-$\ket{211}$ Bohr transitions respectively. The color bar shows corrections on the merger time, rescaled by ratio between the initial cloud mass and the black hole mass $M_c^i/M$. We focus on the regions that $\tau^{\rm sat}_{n\ell m} < 10^7 {\rm yrs} < \tau^{\rm GW}_{n\ell m}$ such that superradiant cloud can form before level transitions. We take $q=1$ as benchmark, and assume that the binaries are initially detected at $0.01$ Hz. We also assume the cloud mass is small enough such that its effects on the orbital dynamics is perturbative, and hence the corrections on the merger time are proportional to the initial cloud mass $M_c^i$.}
    \label{fig:Dt}
\end{figure}

If a transition happens after the binaries passing the LISA band, it still alters the merger time and can be inferred with multiband observations. The merger time of a given orbit is
\be
t_{\rm mgr} = 
\int_{\Omega_i}^{\Omega_{\rm c}} \frac{{\rm d}\Omega}{\dot{\Omega}}\,,
\ee
where $\Omega_i$ is the initial orbital frequency, and $\Omega_c \gg \Omega_i$ can be chosen to be the frequency of the innermost stable circular orbit. In the absence of the cloud, the orbit decay by GWs radiation, and the merger time is given by
\be
t_{\rm mgr}^{\rm vac} =
\frac{5}{256}
\frac{(1+q)^{1/3}}{q}
(GM)^{-5/3} \Omega_i^{-8/3} \,.
\ee
In the presence of the cloud, we have $\dot{\Omega}$ given by Eq.~\eqref{eq:dOmega}, and
\be
t_{\rm mgr}^{\rm cld} =
\int_{\Omega_i}^{\Omega_{\rm c}}
\frac{1}{\gamma_{\rm res}+\gamma_{\rm GW} + \gamma_{\rm ion}} {\rm d}\Omega
.
\ee
When cloud mass is relatively small, the corrections on the merger time from the cloud can be computed perturbatively, i.e., $t_{\rm mg}^{\rm cld} \simeq t_{\rm mg}^{\rm vac} + \Delta t_{\rm mg}^{\rm cld}$ with
\be
\ba\label{eq:Dt}
\Delta t_{\rm mgr}^{\rm cld} \simeq \int_{\Omega_i}^{\Omega_{\rm c}}
-\frac{\gamma_{\rm res}+ \gamma_{\rm ion}}{\gamma_{\rm GW}^2} d\Omega \,.
\ea
\ee
Again, to the leading order, the cloud mass $M_c({\Omega})$ in the above integral can be computed with Eq.~\eqref{eq:Mcapp}. In Fig.~\ref{fig:Dt}, we consider the $\ket{211}$-to-$\ket{100}$ and $\ket{322}$-to-$\ket{211}$ transitions, and compute the corrections on the merger time by integrating Eq.~\eqref{eq:Dt} numerically. We only consider the cases of $M_c^i/M < 10^{-3}$ such that the cloud effects are perturbative.

\section{Detectability of cloud level transitions}\label{sec:detectability}

We now assess how the cloud-induced orbital effects described above can be constrained with milli-Hertz and multiband GW observations. These observations provide two complementary probes: a space-based detector can measure changes in the inspiral phase during the observation, whereas a subsequent ground-based detection can test the coalescence time predicted from the earlier signal. Motivated by the source population in Sec.~\ref{SMBHB}, we consider a nearby equal-mass binary for the milli-Hertz injection-recovery tests, with ${\cal M}_c=30\,M_\odot$, $q=1$, and $D_L=200\,{\rm Mpc}$ observed over $T_{\rm obs}=4\,{\rm yr}$. The injected signal starts at a time before merger $\tau_c=4\,{\rm yr}$, corresponding to an initial frequency of $f_{\rm ini}\simeq 0.016\,{\rm Hz}$. We denote the initial cloud mass fraction by $\varepsilon_c^i\equiv M_c^i/M$, where $M$ is the mass of the black hole hosting the cloud.

\subsection{milli-Hertz observations}\label{sec:milli-Hertz_detection}

For direct observations in the milli-Hertz band, we focus on the $\ket{322}\rightarrow\ket{31-1}$ fine transition identified in Sec.~\ref{sec:para}. We use a TaylorF2 waveform for the vacuum inspiral and include the cloud-induced phase correction in the recovery model. Comparing a vacuum injection with an injection containing a cloud allows us to examine both the behavior of the inference near the vacuum limit and the recovery of nonzero cloud parameters. In each case, the cloud contribution must be inferred jointly with the binary parameters, rather than assessed from the accumulated phase shift alone.

\begin{figure}
\centering
\includegraphics[width=\linewidth]{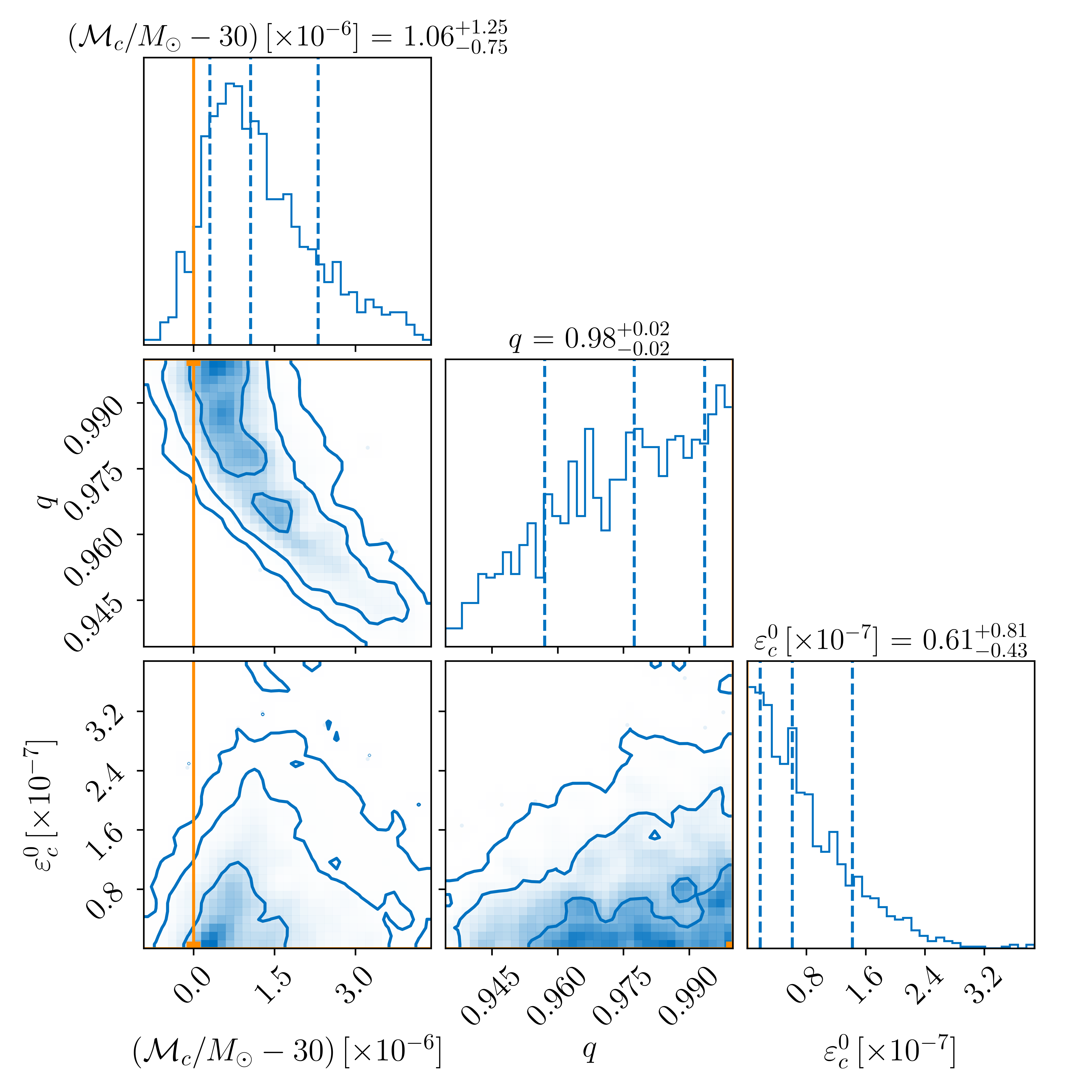}
\caption{Marginalized posterior distributions for a vacuum TaylorF2 injection with ${\cal M}_c=30\,M_\odot$, $q=1$, $D_L=200\,{\rm Mpc}$, and $\varepsilon_c^i=0$, for a four-year LISA observation. The recovery model includes a cloud-induced phase correction. Orange lines mark the injected values; the chirp-mass coordinate is shown relative to its injected values.}
\label{fig:injection_recovery}
\end{figure}

We first inject a vacuum signal, corresponding to $\varepsilon_c^i=0$. Figure~\ref{fig:injection_recovery} shows the posterior distributions for the chirp mass, mass ratio, and cloud-amplitude parameter from nested sampling estimation using \texttt{BILBY} \cite{bilby_paper}. The chirp mass is tightly localized near the injected value, while the cloud amplitude remains consistent with zero, with most of its posterior support at $\varepsilon_c^i\lesssim10^{-6}$. The joint distributions also exhibit correlations between the cloud amplitude and the binary parameters.

\begin{figure}
\centering
\includegraphics[width=\linewidth]{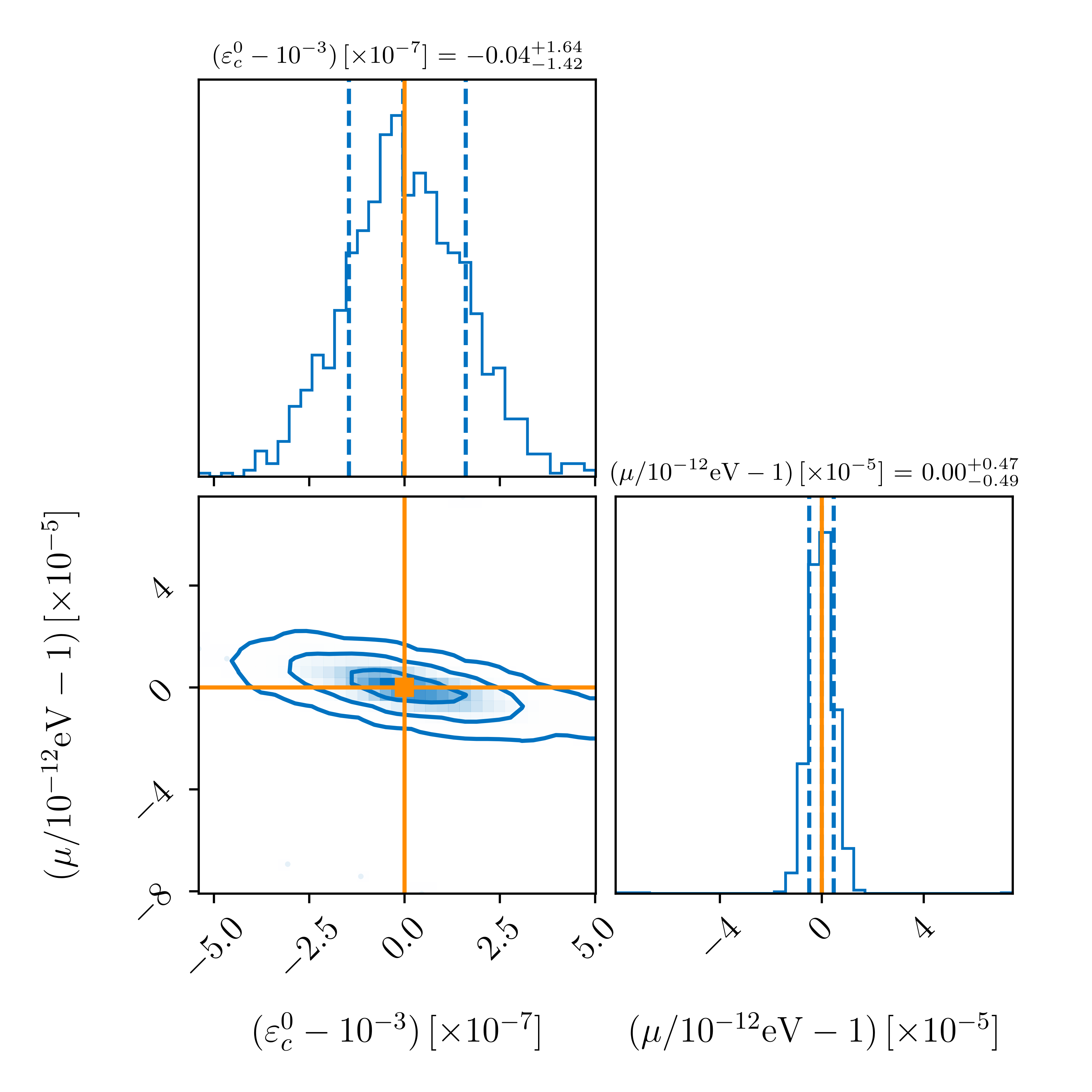}
\caption{Marginalized posterior distributions for an injection containing a cloud undergoing the $\ket{322}$-to-$\ket{31-1}$ transition, with $\mu=1.0\times10^{-12}\,{\rm eV}$ and $\varepsilon_c^i=10^{-3}$. The binary parameters and four-year observation duration are the same as in Fig.~\ref{fig:injection_recovery}. Orange lines mark the injected values; the parameter coordinates are shown relative to those values.}
\label{fig:injection_recovery_boson}
\end{figure}

We next inject a signal containing a cloud with $\mu=1.0\times10^{-12}\,{\rm eV}$ and $\varepsilon_c^i=10^{-3}$, keeping the binary parameters and observation duration unchanged. We present the posterior distribution of $\varepsilon_c^i$ and $\mu$ from nested sampling estimation while fixing other waveform parameters in Fig.~\ref{fig:injection_recovery_boson}. The inferred boson mass and cloud mass fraction are consistent with the injected values, with approximate fractional uncertainties of $0.02\%$ and $0.0005\%$, respectively. The correlation between $\varepsilon_c^i$ and $\mu$ emphasizes the importance of a joint analysis. For this benchmark, the results support the possibility of simultaneously constraining the boson mass and cloud mass fraction through the $\ket{322}\rightarrow\ket{31-1}$ transition. These parameter-recovery uncertainties do not by themselves establish a detection significance, which would require a comparison of the cloud and vacuum hypotheses.

\subsection{Multiband observations}\label{sec:multiband_detection}

Multiband observations provide an indirect probe of cloud evolution that occurs after the space-based observation. If the cloud has a negligible effect on the observed milli-Hertz inspiral, the inferred binary parameters can be used to predict the coalescence time under vacuum evolution. A ground-based measurement of the merger time then tests this prediction. The relevant signal is the cloud-induced correction $\Delta t_{\rm mgr}^{\rm cld}$ in Eq.~\eqref{eq:Dt}, which includes both resonant transitions and ionization. This correction need not be a delay: in the ionization-dominated examples of Fig.~\ref{fig:Dt}, the merger occurs earlier than predicted in vacuum, so $\Delta t_{\rm mgr}^{\rm cld}<0$. We therefore compare the magnitude of the timing shift with the prediction uncertainty. Both quantities must be expressed in the observer frame, consistently with the definition of $\tau_c$ in Sec.~\ref{SMBHB}.

The precision of the space-based prediction depends on the time to coalescence at the start of the observation. Table~\ref{tab:multiband_parameter_estimation} summarizes representative parameter-estimation results for different values of $\tau_c$. For the $\tau_c=10\,{\rm yr}$ case, the reported fractional uncertainty $\Delta\tau_c/\tau_c=2\times10^{-5}$ corresponds to
\begin{equation}
    \Delta\tau_c=(2\times10^{-5})(10\,{\rm yr})
    \simeq6\times10^3\,{\rm s}.
    \label{eq:multiband_timing_precision}
\end{equation}
This uncertainty sets a characteristic scale against which to compare the additional orbital evolution between the space-based observation and the merger.
\begin{table}[!htbp]
\centering
\caption{Representative space-based parameter-estimation results for binaries with different observer-frame times to coalescence $\tau_c$, measured from the start of the observation. The chirp-mass and timing columns give fractional uncertainties; the mass-ratio and luminosity-distance columns give inferred values with their quoted uncertainty ranges.}
\label{tab:multiband_parameter_estimation}
\begin{tabular}{ccccc}
\hline
$\tau_c$
& $\Delta\mathcal{M}_c/\mathcal{M}_c$
& $q$
& $D_L\ [{\rm Mpc}]$
& $\Delta\tau_c/\tau_c$ \\
\hline
$1\ {\rm yr}$
& $6\times10^{-6}$
& $0.8^{+0.1}_{-0.2}$
& $101^{+20}_{-20}$
& $4\times10^{-7}$ \\

$4\ {\rm yr}$
& $3\times10^{-6}$
& $0.8^{+0.2}_{-0.1}$
& $212^{+40}_{-42}$
& $2\times10^{-7}$ \\

$10\ {\rm yr}$
& $3\times10^{-5}$
& $0.6^{+0.2}_{-0.3}$
& $181^{+34}_{-35}$
& $2\times10^{-5}$ \\

$20\ {\rm yr}$
& $3\times10^{-5}$
& $0.6^{+0.3}_{-0.3}$
& $155^{+29}_{-31}$
& $2\times10^{-5}$ \\
\hline
\end{tabular}
\end{table}

In the perturbative regime of Eqs.~\eqref{eq:Dt} and~\eqref{eq:Mcapp}, the timing correction is linear in the initial cloud mass. We write
\begin{equation}
    |\Delta t_{\rm mgr}^{\rm cld}|
    \simeq K_{\rm cld}\frac{M_c^i}{M},
    \label{eq:cloud_timing_response}
\end{equation}
where the response coefficient $K_{\rm cld}$ depends on the initial cloud state, boson mass, binary parameters, and interval over which the orbital evolution is compared. Equating this shift to the prediction uncertainty gives the indicative sensitivity
\begin{equation}
    \begin{aligned}
        \left(\frac{M_c^i}{M}\right)_{\rm sens}
        &\sim\frac{\Delta\tau_c}{K_{\rm cld}}\\
        &\sim2\times10^{-5}
        \left(\frac{\Delta\tau_c}{6\times10^3\,{\rm s}}\right)
        \left(\frac{K_{\rm cld}}{2.8\times10^8\,{\rm s}}\right)^{-1}.
    \end{aligned}
    \label{eq:multiband_cloud_sensitivity}
\end{equation}
The coefficient $2.8\times10^8\,{\rm s}$ is used here as an illustrative normalization. Associating this sensitivity with a particular boson mass or either Bohr transition requires evaluating $K_{\rm cld}$ for the corresponding system using Eq.~\eqref{eq:Dt}; it is not a universal cloud-mass threshold.

This estimate assumes that the ground-based timing uncertainty and errors in the waveform model are negligible compared with the uncertainty in the space-based prediction. It characterizes the sensitivity of a consistency test, not a detection significance. A timing discrepancy constrains the accumulated nonvacuum orbital evolution and does not uniquely identify a cloud or a particular level transition. A cloud interpretation would therefore require a joint analysis of the two observations and comparison with alternative explanations. The two observing strategies probe complementary stages of cloud-binary evolution; their sensitivities do not refer to the same transition under identical observing conditions.

\section{Conclusion and Discussion}\label{sec:conclusion}

We have investigated how superradiant cloud level transitions can be probed with milli-Hertz and multiband GW observations of SMBHBs. The coupled evolution of the cloud and the orbit gives rise to two complementary observables: a correction to the inspiral phase measured by LISA and a departure from the vacuum coalescence time predicted from the early inspiral. Under the adopted cloud formation and survival conditions, the $\ket{322}\rightarrow\ket{31-1}$ fine transition provides a target for direct milli-Hertz observations, while the $\ket{211}\rightarrow\ket{100}$ and $\ket{322}\rightarrow\ket{211}$ Bohr transitions motivate an indirect multiband timing test.

For our fiducial population and selection models, a four-year LISA observation yields approximately $4.6$ detectable binaries above an SNR threshold of $8$. Requiring a ground-based detection and a merger within $20\,{\rm yr}$ of the start of the LISA observation reduces the expected multiband sample to approximately $1.4$. Lowering the LISA threshold to $5$, as an illustrative ground-informed search scenario, increases the multiband expectation to approximately $6.7$. These counts characterize the available binary sample, not the number of detectable cloud transitions. The latter additionally depends on the abundance and survival of clouds and on the observability of their orbital effects.

The fine-transition dynamics show that cloud depletion can begin before the nominal resonance, making it important to evolve the cloud mass together with the orbital frequency. For the representative equal-mass binary with ${\cal M}_c=30\,M_\odot$ at $200\,{\rm Mpc}$, observed for four years, the vacuum injection yields a cloud-amplitude posterior consistent with zero, with most support at $M_c^i/M\lesssim10^{-6}$. An injection with $\mu=10^{-12}\,{\rm eV}$ and $M_c^i/M=10^{-3}$ recovers both cloud parameters when the remaining waveform parameters are fixed. These examples illustrate the information encoded in the cloud-induced phase correction; they do not establish a general detection threshold, a fully marginalized measurement precision, or a detection significance.

For cloud evolution occurring after the space-based observation, multiband timing tests the vacuum coalescence-time prediction, provided that cloud effects on the earlier LISA signal are negligible. This test must include ionization as well as resonant transfer: in the ionization-dominated examples considered here, the merger is advanced rather than delayed. In the perturbative regime, the timing correction scales linearly with the initial cloud mass. For the $\tau_c=10\,{\rm yr}$ example, a prediction uncertainty of approximately $6\times10^3\,{\rm s}$ corresponds to an indicative sensitivity of $M_c^i/M\sim2\times10^{-5}$ for the illustrative response coefficient $K_{\rm cld}=2.8\times10^8\,{\rm s}$. This sensitivity must be evaluated separately for each binary and cloud configuration; it is not a universal threshold for either Bohr transition. Moreover, a timing discrepancy constrains accumulated nonvacuum orbital evolution and does not uniquely identify a boson cloud.

The interpretation of these results depends on the history of the cloud, not only on the location of a resonance. The conditions in Sec.~\ref{sec:para}, evaluated at a fiducial age of $10^7\,{\rm yr}$, identify possible cloud configurations but do not determine their abundance among LISA sources. Initial black hole spins, binary ages, and prior orbital evolution must be connected to the cloud state and remaining mass at the start of the observation. The pre-resonance depletion found here illustrates why an allowed transition need not produce an observable signal. Conversely, a null result can reflect the absence or prior depletion of a cloud, and does not by itself exclude a boson mass. A self-consistent population forecast must also revisit the vacuum selection used in Sec.~\ref{SMBHB} whenever cloud backreaction appreciably changes the residence time in the LISA band or the time to merger. The cloud-hosting fraction and the cloud-modified detection window are therefore separate ingredients in translating single-source sensitivities into event yields.

The two observables also retain different information about the orbital perturbation. The phase correction in Eq.~\eqref{eq:DPhi} can carry a frequency-dependent signature, but constant and linear-in-frequency terms are degenerate with $\phi_c$ and $t_c$, while the remaining terms can correlate with the binary masses. A large accumulated dephasing is thus not equivalent to a distinguishable cloud signal. This motivates joint inference of binary and cloud parameters, with consistent waveform and detector-response modeling and explicit comparison of cloud, vacuum, and alternative nonvacuum models. By contrast, the timing shift in Eq.~\eqref{eq:Dt} integrates the subsequent evolution into a single offset, which cannot reconstruct the cloud history uniquely. For a single binary, combining phase and timing information also requires propagating the cloud state and mass consistently between observing epochs: depletion during an earlier transition may weaken a later timing signature. Complementarity should therefore not be taken to imply that both signatures must be observable in the same system.

The dynamical approximations impose another important qualification. The linear phase and timing corrections require $|\gamma_{\rm res}+\gamma_{\rm ion}|\ll\gamma_{\rm GW}$ throughout the relevant evolution; a small mass fraction $M_c^i/M$ alone does not ensure weak orbital backreaction near resonance. The floating-orbit condition and the temporary outspiral discussed above accordingly motivate waveform calculations based on the coupled dynamics, rather than extrapolation of the perturbative sensitivity scalings. Moreover, the reduced transition rates assume $|\eta_{ba}|\ll|\Gamma_b|$. Outside this regime, the dynamics of the target state must be retained rather than replaced by its quasistationary occupation. Extending the calculation beyond quasicircular, spin-aligned binaries would likewise require revisiting the tidal couplings and the sequence of accessible transitions. These extensions are necessary before the benchmark results can be applied to a generic binary population.

These considerations suggest several priorities for future observations and analysis. Ground-informed archival searches can enlarge the multiband sample, but the additional low-SNR sources will not necessarily retain the cloud-parameter precision of the benchmark analyzed here. The effectiveness of this strategy must be established with injection-recovery studies and a consistent treatment of selection effects. A sufficiently sensitive deci-Hertz observation could instead resolve part of the evolution that the timing test measures only through its integrated effect, providing additional information on the relative roles of resonant transfer and ionization. Its quantitative reach requires a dedicated forecast. Milli-Hertz phase measurements and multiband timing therefore offer complementary routes to probing ultralight bosons: they test both the cloud's imprint during an observed inspiral and its accumulated orbital influence between separately observed stages of a binary.

\section*{ACKNOWLEDGMENTS}

J. Z. is supported by the scientific research starting grants from the University of Chinese Academy of Sciences (Grant No.~118900M061), the Fundamental Research Funds for the Central Universities (Grants No.~E2EG6602X2 and No.~E2ET0209X2), the National Natural Science Foundation of China (NSFC) under Grant No.~12147103 and National Key Research and Development Program of China (Project No. 2025YFE0217300). A.C. is supported by the China Postdoctoral Science Foundation under Grant No. 2025M773325, and the NSFC under Grant No. E414660101 and 12147103.

\bibliography{references}

\begin{thebibliography}{52}%
\makeatletter
\providecommand \@ifxundefined [1]{%
 \@ifx{#1\undefined}
}%
\providecommand \@ifnum [1]{%
 \ifnum #1\expandafter \@firstoftwo
 \else \expandafter \@secondoftwo
 \fi
}%
\providecommand \@ifx [1]{%
 \ifx #1\expandafter \@firstoftwo
 \else \expandafter \@secondoftwo
 \fi
}%
\providecommand \natexlab [1]{#1}%
\providecommand \enquote  [1]{``#1''}%
\providecommand \bibnamefont  [1]{#1}%
\providecommand \bibfnamefont [1]{#1}%
\providecommand \citenamefont [1]{#1}%
\providecommand \href@noop [0]{\@secondoftwo}%
\providecommand \href [0]{\begingroup \@sanitize@url \@href}%
\providecommand \@href[1]{\@@startlink{#1}\@@href}%
\providecommand \@@href[1]{\endgroup#1\@@endlink}%
\providecommand \@sanitize@url [0]{\catcode `\\12\catcode `\$12\catcode
  `\&12\catcode `\#12\catcode `\^12\catcode `\_12\catcode `\%12\relax}%
\providecommand \@@startlink[1]{}%
\providecommand \@@endlink[0]{}%
\providecommand \url  [0]{\begingroup\@sanitize@url \@url }%
\providecommand \@url [1]{\endgroup\@href {#1}{\urlprefix }}%
\providecommand \urlprefix  [0]{URL }%
\providecommand \Eprint [0]{\href }%
\providecommand \doibase [0]{https://doi.org/}%
\providecommand \selectlanguage [0]{\@gobble}%
\providecommand \bibinfo  [0]{\@secondoftwo}%
\providecommand \bibfield  [0]{\@secondoftwo}%
\providecommand \translation [1]{[#1]}%
\providecommand \BibitemOpen [0]{}%
\providecommand \bibitemStop [0]{}%
\providecommand \bibitemNoStop [0]{.\EOS\space}%
\providecommand \EOS [0]{\spacefactor3000\relax}%
\providecommand \BibitemShut  [1]{\csname bibitem#1\endcsname}%
\let\auto@bib@innerbib\@empty
\bibitem [{\citenamefont {Arvanitaki}\ \emph {et~al.}(2010)\citenamefont
  {Arvanitaki}, \citenamefont {Dimopoulos}, \citenamefont {Dubovsky},
  \citenamefont {Kaloper},\ and\ \citenamefont
  {March-Russell}}]{Arvanitaki:2009fg}%
  \BibitemOpen
  \bibfield  {author} {\bibinfo {author} {\bibfnamefont {A.}~\bibnamefont
  {Arvanitaki}}, \bibinfo {author} {\bibfnamefont {S.}~\bibnamefont
  {Dimopoulos}}, \bibinfo {author} {\bibfnamefont {S.}~\bibnamefont
  {Dubovsky}}, \bibinfo {author} {\bibfnamefont {N.}~\bibnamefont {Kaloper}},\
  and\ \bibinfo {author} {\bibfnamefont {J.}~\bibnamefont {March-Russell}},\
  }\bibfield  {title} {\bibinfo {title} {{String Axiverse}},\ }\href
  {https://doi.org/10.1103/PhysRevD.81.123530} {\bibfield  {journal} {\bibinfo
  {journal} {Phys. Rev. D}\ }\textbf {\bibinfo {volume} {81}},\ \bibinfo
  {pages} {123530} (\bibinfo {year} {2010})},\ \Eprint
  {https://arxiv.org/abs/0905.4720} {arXiv:0905.4720 [hep-th]} \BibitemShut
  {NoStop}%
\bibitem [{\citenamefont {Hui}\ \emph {et~al.}(2017)\citenamefont {Hui},
  \citenamefont {Ostriker}, \citenamefont {Tremaine},\ and\ \citenamefont
  {Witten}}]{Hui:2016ltb}%
  \BibitemOpen
  \bibfield  {author} {\bibinfo {author} {\bibfnamefont {L.}~\bibnamefont
  {Hui}}, \bibinfo {author} {\bibfnamefont {J.~P.}\ \bibnamefont {Ostriker}},
  \bibinfo {author} {\bibfnamefont {S.}~\bibnamefont {Tremaine}},\ and\
  \bibinfo {author} {\bibfnamefont {E.}~\bibnamefont {Witten}},\ }\bibfield
  {title} {\bibinfo {title} {{Ultralight scalars as cosmological dark
  matter}},\ }\href {https://doi.org/10.1103/PhysRevD.95.043541} {\bibfield
  {journal} {\bibinfo  {journal} {Phys. Rev. D}\ }\textbf {\bibinfo {volume}
  {95}},\ \bibinfo {pages} {043541} (\bibinfo {year} {2017})},\ \Eprint
  {https://arxiv.org/abs/1610.08297} {arXiv:1610.08297 [astro-ph.CO]}
  \BibitemShut {NoStop}%
\bibitem [{\citenamefont {Brito}\ \emph {et~al.}(2015)\citenamefont {Brito},
  \citenamefont {Cardoso},\ and\ \citenamefont {Pani}}]{Brito:2015oca}%
  \BibitemOpen
  \bibfield  {author} {\bibinfo {author} {\bibfnamefont {R.}~\bibnamefont
  {Brito}}, \bibinfo {author} {\bibfnamefont {V.}~\bibnamefont {Cardoso}},\
  and\ \bibinfo {author} {\bibfnamefont {P.}~\bibnamefont {Pani}},\ }\bibfield
  {title} {\bibinfo {title} {{Superradiance}: {New Frontiers in Black Hole
  Physics}},\ }\href {https://doi.org/10.1007/978-3-319-19000-6} {\bibfield
  {journal} {\bibinfo  {journal} {Lect. Notes Phys.}\ }\textbf {\bibinfo
  {volume} {906}},\ \bibinfo {pages} {pp.1} (\bibinfo {year} {2015})},\ \Eprint
  {https://arxiv.org/abs/1501.06570} {arXiv:1501.06570 [gr-qc]} \BibitemShut
  {NoStop}%
\bibitem [{\citenamefont {Arvanitaki}\ \emph {et~al.}(2015)\citenamefont
  {Arvanitaki}, \citenamefont {Baryakhtar},\ and\ \citenamefont
  {Huang}}]{Arvanitaki:2014wva}%
  \BibitemOpen
  \bibfield  {author} {\bibinfo {author} {\bibfnamefont {A.}~\bibnamefont
  {Arvanitaki}}, \bibinfo {author} {\bibfnamefont {M.}~\bibnamefont
  {Baryakhtar}},\ and\ \bibinfo {author} {\bibfnamefont {X.}~\bibnamefont
  {Huang}},\ }\bibfield  {title} {\bibinfo {title} {{Discovering the QCD Axion
  with Black Holes and Gravitational Waves}},\ }\href
  {https://doi.org/10.1103/PhysRevD.91.084011} {\bibfield  {journal} {\bibinfo
  {journal} {Phys. Rev. D}\ }\textbf {\bibinfo {volume} {91}},\ \bibinfo
  {pages} {084011} (\bibinfo {year} {2015})},\ \Eprint
  {https://arxiv.org/abs/1411.2263} {arXiv:1411.2263 [hep-ph]} \BibitemShut
  {NoStop}%
\bibitem [{\citenamefont {Brito}\ \emph
  {et~al.}(2017{\natexlab{a}})\citenamefont {Brito}, \citenamefont {Ghosh},
  \citenamefont {Barausse}, \citenamefont {Berti}, \citenamefont {Cardoso},
  \citenamefont {Dvorkin}, \citenamefont {Klein},\ and\ \citenamefont
  {Pani}}]{Brito:2017wnc}%
  \BibitemOpen
  \bibfield  {author} {\bibinfo {author} {\bibfnamefont {R.}~\bibnamefont
  {Brito}}, \bibinfo {author} {\bibfnamefont {S.}~\bibnamefont {Ghosh}},
  \bibinfo {author} {\bibfnamefont {E.}~\bibnamefont {Barausse}}, \bibinfo
  {author} {\bibfnamefont {E.}~\bibnamefont {Berti}}, \bibinfo {author}
  {\bibfnamefont {V.}~\bibnamefont {Cardoso}}, \bibinfo {author} {\bibfnamefont
  {I.}~\bibnamefont {Dvorkin}}, \bibinfo {author} {\bibfnamefont
  {A.}~\bibnamefont {Klein}},\ and\ \bibinfo {author} {\bibfnamefont
  {P.}~\bibnamefont {Pani}},\ }\bibfield  {title} {\bibinfo {title}
  {{Stochastic and resolvable gravitational waves from ultralight bosons}},\
  }\href {https://doi.org/10.1103/PhysRevLett.119.131101} {\bibfield  {journal}
  {\bibinfo  {journal} {Phys. Rev. Lett.}\ }\textbf {\bibinfo {volume} {119}},\
  \bibinfo {pages} {131101} (\bibinfo {year} {2017}{\natexlab{a}})},\ \Eprint
  {https://arxiv.org/abs/1706.05097} {arXiv:1706.05097 [gr-qc]} \BibitemShut
  {NoStop}%
\bibitem [{\citenamefont {Brito}\ \emph
  {et~al.}(2017{\natexlab{b}})\citenamefont {Brito}, \citenamefont {Ghosh},
  \citenamefont {Barausse}, \citenamefont {Berti}, \citenamefont {Cardoso},
  \citenamefont {Dvorkin}, \citenamefont {Klein},\ and\ \citenamefont
  {Pani}}]{Brito:2017zvb}%
  \BibitemOpen
  \bibfield  {author} {\bibinfo {author} {\bibfnamefont {R.}~\bibnamefont
  {Brito}}, \bibinfo {author} {\bibfnamefont {S.}~\bibnamefont {Ghosh}},
  \bibinfo {author} {\bibfnamefont {E.}~\bibnamefont {Barausse}}, \bibinfo
  {author} {\bibfnamefont {E.}~\bibnamefont {Berti}}, \bibinfo {author}
  {\bibfnamefont {V.}~\bibnamefont {Cardoso}}, \bibinfo {author} {\bibfnamefont
  {I.}~\bibnamefont {Dvorkin}}, \bibinfo {author} {\bibfnamefont
  {A.}~\bibnamefont {Klein}},\ and\ \bibinfo {author} {\bibfnamefont
  {P.}~\bibnamefont {Pani}},\ }\bibfield  {title} {\bibinfo {title}
  {{Gravitational wave searches for ultralight bosons with LIGO and LISA}},\
  }\href {https://doi.org/10.1103/PhysRevD.96.064050} {\bibfield  {journal}
  {\bibinfo  {journal} {Phys. Rev. D}\ }\textbf {\bibinfo {volume} {96}},\
  \bibinfo {pages} {064050} (\bibinfo {year} {2017}{\natexlab{b}})},\ \Eprint
  {https://arxiv.org/abs/1706.06311} {arXiv:1706.06311 [gr-qc]} \BibitemShut
  {NoStop}%
\bibitem [{\citenamefont {Tsukada}\ \emph {et~al.}(2019)\citenamefont
  {Tsukada}, \citenamefont {Callister}, \citenamefont {Matas},\ and\
  \citenamefont {Meyers}}]{Tsukada:2018mbp}%
  \BibitemOpen
  \bibfield  {author} {\bibinfo {author} {\bibfnamefont {L.}~\bibnamefont
  {Tsukada}}, \bibinfo {author} {\bibfnamefont {T.}~\bibnamefont {Callister}},
  \bibinfo {author} {\bibfnamefont {A.}~\bibnamefont {Matas}},\ and\ \bibinfo
  {author} {\bibfnamefont {P.}~\bibnamefont {Meyers}},\ }\bibfield  {title}
  {\bibinfo {title} {{First search for a stochastic gravitational-wave
  background from ultralight bosons}},\ }\href
  {https://doi.org/10.1103/PhysRevD.99.103015} {\bibfield  {journal} {\bibinfo
  {journal} {Phys. Rev. D}\ }\textbf {\bibinfo {volume} {99}},\ \bibinfo
  {pages} {103015} (\bibinfo {year} {2019})},\ \Eprint
  {https://arxiv.org/abs/1812.09622} {arXiv:1812.09622 [astro-ph.HE]}
  \BibitemShut {NoStop}%
\bibitem [{\citenamefont {Isi}\ \emph {et~al.}(2019)\citenamefont {Isi},
  \citenamefont {Sun}, \citenamefont {Brito},\ and\ \citenamefont
  {Melatos}}]{Isi:2018pzk}%
  \BibitemOpen
  \bibfield  {author} {\bibinfo {author} {\bibfnamefont {M.}~\bibnamefont
  {Isi}}, \bibinfo {author} {\bibfnamefont {L.}~\bibnamefont {Sun}}, \bibinfo
  {author} {\bibfnamefont {R.}~\bibnamefont {Brito}},\ and\ \bibinfo {author}
  {\bibfnamefont {A.}~\bibnamefont {Melatos}},\ }\bibfield  {title} {\bibinfo
  {title} {{Directed searches for gravitational waves from ultralight
  bosons}},\ }\href {https://doi.org/10.1103/PhysRevD.99.084042} {\bibfield
  {journal} {\bibinfo  {journal} {Phys. Rev. D}\ }\textbf {\bibinfo {volume}
  {99}},\ \bibinfo {pages} {084042} (\bibinfo {year} {2019})},\ \bibinfo {note}
  {[Erratum: Phys.Rev.D 102, 049901 (2020)]},\ \Eprint
  {https://arxiv.org/abs/1810.03812} {arXiv:1810.03812 [gr-qc]} \BibitemShut
  {NoStop}%
\bibitem [{\citenamefont {Palomba}\ \emph {et~al.}(2019)\citenamefont {Palomba}
  \emph {et~al.}}]{Palomba:2019vxe}%
  \BibitemOpen
  \bibfield  {author} {\bibinfo {author} {\bibfnamefont {C.}~\bibnamefont
  {Palomba}} \emph {et~al.},\ }\bibfield  {title} {\bibinfo {title} {{Direct
  constraints on ultra-light boson mass from searches for continuous
  gravitational waves}},\ }\href
  {https://doi.org/10.1103/PhysRevLett.123.171101} {\bibfield  {journal}
  {\bibinfo  {journal} {Phys. Rev. Lett.}\ }\textbf {\bibinfo {volume} {123}},\
  \bibinfo {pages} {171101} (\bibinfo {year} {2019})},\ \Eprint
  {https://arxiv.org/abs/1909.08854} {arXiv:1909.08854 [astro-ph.HE]}
  \BibitemShut {NoStop}%
\bibitem [{\citenamefont {Tsukada}\ \emph {et~al.}(2021)\citenamefont
  {Tsukada}, \citenamefont {Brito}, \citenamefont {East},\ and\ \citenamefont
  {Siemonsen}}]{Tsukada:2020lgt}%
  \BibitemOpen
  \bibfield  {author} {\bibinfo {author} {\bibfnamefont {L.}~\bibnamefont
  {Tsukada}}, \bibinfo {author} {\bibfnamefont {R.}~\bibnamefont {Brito}},
  \bibinfo {author} {\bibfnamefont {W.~E.}\ \bibnamefont {East}},\ and\
  \bibinfo {author} {\bibfnamefont {N.}~\bibnamefont {Siemonsen}},\ }\bibfield
  {title} {\bibinfo {title} {{Modeling and searching for a stochastic
  gravitational-wave background from ultralight vector bosons}},\ }\href
  {https://doi.org/10.1103/PhysRevD.103.083005} {\bibfield  {journal} {\bibinfo
   {journal} {Phys. Rev. D}\ }\textbf {\bibinfo {volume} {103}},\ \bibinfo
  {pages} {083005} (\bibinfo {year} {2021})},\ \Eprint
  {https://arxiv.org/abs/2011.06995} {arXiv:2011.06995 [astro-ph.HE]}
  \BibitemShut {NoStop}%
\bibitem [{\citenamefont {Ng}\ \emph {et~al.}(2021)\citenamefont {Ng},
  \citenamefont {Vitale}, \citenamefont {Hannuksela},\ and\ \citenamefont
  {Li}}]{Ng:2020ruv}%
  \BibitemOpen
  \bibfield  {author} {\bibinfo {author} {\bibfnamefont {K.~K.~Y.}\
  \bibnamefont {Ng}}, \bibinfo {author} {\bibfnamefont {S.}~\bibnamefont
  {Vitale}}, \bibinfo {author} {\bibfnamefont {O.~A.}\ \bibnamefont
  {Hannuksela}},\ and\ \bibinfo {author} {\bibfnamefont {T.~G.~F.}\
  \bibnamefont {Li}},\ }\bibfield  {title} {\bibinfo {title} {{Constraints on
  Ultralight Scalar Bosons within Black Hole Spin Measurements from the
  LIGO-Virgo GWTC-2}},\ }\href {https://doi.org/10.1103/PhysRevLett.126.151102}
  {\bibfield  {journal} {\bibinfo  {journal} {Phys. Rev. Lett.}\ }\textbf
  {\bibinfo {volume} {126}},\ \bibinfo {pages} {151102} (\bibinfo {year}
  {2021})},\ \Eprint {https://arxiv.org/abs/2011.06010} {arXiv:2011.06010
  [gr-qc]} \BibitemShut {NoStop}%
\bibitem [{\citenamefont {Abbott}\ \emph {et~al.}(2022)\citenamefont {Abbott}
  \emph {et~al.}}]{LIGOScientific:2021rnv}%
  \BibitemOpen
  \bibfield  {author} {\bibinfo {author} {\bibfnamefont {R.}~\bibnamefont
  {Abbott}} \emph {et~al.} (\bibinfo {collaboration} {LIGO Scientific, Virgo,
  KAGRA}),\ }\bibfield  {title} {\bibinfo {title} {{All-sky search for
  gravitational wave emission from scalar boson clouds around spinning black
  holes in LIGO O3 data}},\ }\href
  {https://doi.org/10.1103/PhysRevD.105.102001} {\bibfield  {journal} {\bibinfo
   {journal} {Phys. Rev. D}\ }\textbf {\bibinfo {volume} {105}},\ \bibinfo
  {pages} {102001} (\bibinfo {year} {2022})},\ \Eprint
  {https://arxiv.org/abs/2111.15507} {arXiv:2111.15507 [astro-ph.HE]}
  \BibitemShut {NoStop}%
\bibitem [{\citenamefont {Yuan}\ \emph {et~al.}(2022)\citenamefont {Yuan},
  \citenamefont {Jiang},\ and\ \citenamefont {Huang}}]{Yuan:2022bem}%
  \BibitemOpen
  \bibfield  {author} {\bibinfo {author} {\bibfnamefont {C.}~\bibnamefont
  {Yuan}}, \bibinfo {author} {\bibfnamefont {Y.}~\bibnamefont {Jiang}},\ and\
  \bibinfo {author} {\bibfnamefont {Q.-G.}\ \bibnamefont {Huang}},\ }\bibfield
  {title} {\bibinfo {title} {{Constraints on an ultralight scalar boson from
  Advanced LIGO and Advanced Virgo\textquoteright{}s first three observing runs
  using the stochastic gravitational-wave background}},\ }\href
  {https://doi.org/10.1103/PhysRevD.106.023020} {\bibfield  {journal} {\bibinfo
   {journal} {Phys. Rev. D}\ }\textbf {\bibinfo {volume} {106}},\ \bibinfo
  {pages} {023020} (\bibinfo {year} {2022})},\ \Eprint
  {https://arxiv.org/abs/2204.03482} {arXiv:2204.03482 [astro-ph.CO]}
  \BibitemShut {NoStop}%
\bibitem [{\citenamefont {Aurrekoetxea}\ \emph
  {et~al.}(2024{\natexlab{a}})\citenamefont {Aurrekoetxea}, \citenamefont
  {Clough}, \citenamefont {Bamber},\ and\ \citenamefont
  {Ferreira}}]{Aurrekoetxea:2023jwk}%
  \BibitemOpen
  \bibfield  {author} {\bibinfo {author} {\bibfnamefont {J.~C.}\ \bibnamefont
  {Aurrekoetxea}}, \bibinfo {author} {\bibfnamefont {K.}~\bibnamefont
  {Clough}}, \bibinfo {author} {\bibfnamefont {J.}~\bibnamefont {Bamber}},\
  and\ \bibinfo {author} {\bibfnamefont {P.~G.}\ \bibnamefont {Ferreira}},\
  }\bibfield  {title} {\bibinfo {title} {{Effect of Wave Dark Matter on Equal
  Mass Black Hole Mergers}},\ }\href
  {https://doi.org/10.1103/PhysRevLett.132.211401} {\bibfield  {journal}
  {\bibinfo  {journal} {Phys. Rev. Lett.}\ }\textbf {\bibinfo {volume} {132}},\
  \bibinfo {pages} {211401} (\bibinfo {year} {2024}{\natexlab{a}})},\ \Eprint
  {https://arxiv.org/abs/2311.18156} {arXiv:2311.18156 [gr-qc]} \BibitemShut
  {NoStop}%
\bibitem [{\citenamefont {Aurrekoetxea}\ \emph
  {et~al.}(2024{\natexlab{b}})\citenamefont {Aurrekoetxea}, \citenamefont
  {Marsden}, \citenamefont {Clough},\ and\ \citenamefont
  {Ferreira}}]{Aurrekoetxea:2024cqd}%
  \BibitemOpen
  \bibfield  {author} {\bibinfo {author} {\bibfnamefont {J.~C.}\ \bibnamefont
  {Aurrekoetxea}}, \bibinfo {author} {\bibfnamefont {J.}~\bibnamefont
  {Marsden}}, \bibinfo {author} {\bibfnamefont {K.}~\bibnamefont {Clough}},\
  and\ \bibinfo {author} {\bibfnamefont {P.~G.}\ \bibnamefont {Ferreira}},\
  }\bibfield  {title} {\bibinfo {title} {{Self-interacting scalar dark matter
  around binary black holes}},\ }\href
  {https://doi.org/10.1103/PhysRevD.110.083011} {\bibfield  {journal} {\bibinfo
   {journal} {Phys. Rev. D}\ }\textbf {\bibinfo {volume} {110}},\ \bibinfo
  {pages} {083011} (\bibinfo {year} {2024}{\natexlab{b}})},\ \Eprint
  {https://arxiv.org/abs/2409.01937} {arXiv:2409.01937 [gr-qc]} \BibitemShut
  {NoStop}%
\bibitem [{\citenamefont {Miller}(2025)}]{Miller:2025yyx}%
  \BibitemOpen
  \bibfield  {author} {\bibinfo {author} {\bibfnamefont {A.~L.}\ \bibnamefont
  {Miller}},\ }\bibfield  {title} {\bibinfo {title} {{Gravitational wave probes
  of particle dark matter: a review}},\ }\href@noop {} {\  (\bibinfo {year}
  {2025})},\ \Eprint {https://arxiv.org/abs/2503.02607} {arXiv:2503.02607
  [astro-ph.HE]} \BibitemShut {NoStop}%
\bibitem [{\citenamefont {Baumann}\ \emph {et~al.}(2019)\citenamefont
  {Baumann}, \citenamefont {Chia},\ and\ \citenamefont
  {Porto}}]{Baumann:2018vus}%
  \BibitemOpen
  \bibfield  {author} {\bibinfo {author} {\bibfnamefont {D.}~\bibnamefont
  {Baumann}}, \bibinfo {author} {\bibfnamefont {H.~S.}\ \bibnamefont {Chia}},\
  and\ \bibinfo {author} {\bibfnamefont {R.~A.}\ \bibnamefont {Porto}},\
  }\bibfield  {title} {\bibinfo {title} {{Probing Ultralight Bosons with Binary
  Black Holes}},\ }\href {https://doi.org/10.1103/PhysRevD.99.044001}
  {\bibfield  {journal} {\bibinfo  {journal} {Phys. Rev. D}\ }\textbf {\bibinfo
  {volume} {99}},\ \bibinfo {pages} {044001} (\bibinfo {year} {2019})},\
  \Eprint {https://arxiv.org/abs/1804.03208} {arXiv:1804.03208 [gr-qc]}
  \BibitemShut {NoStop}%
\bibitem [{\citenamefont {Baumann}\ \emph {et~al.}(2020)\citenamefont
  {Baumann}, \citenamefont {Chia}, \citenamefont {Porto},\ and\ \citenamefont
  {Stout}}]{Baumann:2019ztm}%
  \BibitemOpen
  \bibfield  {author} {\bibinfo {author} {\bibfnamefont {D.}~\bibnamefont
  {Baumann}}, \bibinfo {author} {\bibfnamefont {H.~S.}\ \bibnamefont {Chia}},
  \bibinfo {author} {\bibfnamefont {R.~A.}\ \bibnamefont {Porto}},\ and\
  \bibinfo {author} {\bibfnamefont {J.}~\bibnamefont {Stout}},\ }\bibfield
  {title} {\bibinfo {title} {{Gravitational Collider Physics}},\ }\href
  {https://doi.org/10.1103/PhysRevD.101.083019} {\bibfield  {journal} {\bibinfo
   {journal} {Phys. Rev. D}\ }\textbf {\bibinfo {volume} {101}},\ \bibinfo
  {pages} {083019} (\bibinfo {year} {2020})},\ \Eprint
  {https://arxiv.org/abs/1912.04932} {arXiv:1912.04932 [gr-qc]} \BibitemShut
  {NoStop}%
\bibitem [{\citenamefont {Berti}\ \emph {et~al.}(2019)\citenamefont {Berti},
  \citenamefont {Brito}, \citenamefont {Macedo}, \citenamefont {Raposo},\ and\
  \citenamefont {Rosa}}]{Berti:2019wnn}%
  \BibitemOpen
  \bibfield  {author} {\bibinfo {author} {\bibfnamefont {E.}~\bibnamefont
  {Berti}}, \bibinfo {author} {\bibfnamefont {R.}~\bibnamefont {Brito}},
  \bibinfo {author} {\bibfnamefont {C.~F.~B.}\ \bibnamefont {Macedo}}, \bibinfo
  {author} {\bibfnamefont {G.}~\bibnamefont {Raposo}},\ and\ \bibinfo {author}
  {\bibfnamefont {J.~L.}\ \bibnamefont {Rosa}},\ }\bibfield  {title} {\bibinfo
  {title} {{Ultralight boson cloud depletion in binary systems}},\ }\href
  {https://doi.org/10.1103/PhysRevD.99.104039} {\bibfield  {journal} {\bibinfo
  {journal} {Phys. Rev. D}\ }\textbf {\bibinfo {volume} {99}},\ \bibinfo
  {pages} {104039} (\bibinfo {year} {2019})},\ \Eprint
  {https://arxiv.org/abs/1904.03131} {arXiv:1904.03131 [gr-qc]} \BibitemShut
  {NoStop}%
\bibitem [{\citenamefont {Baumann}\ \emph
  {et~al.}(2022{\natexlab{a}})\citenamefont {Baumann}, \citenamefont {Bertone},
  \citenamefont {Stout},\ and\ \citenamefont {Tomaselli}}]{Baumann:2021fkf}%
  \BibitemOpen
  \bibfield  {author} {\bibinfo {author} {\bibfnamefont {D.}~\bibnamefont
  {Baumann}}, \bibinfo {author} {\bibfnamefont {G.}~\bibnamefont {Bertone}},
  \bibinfo {author} {\bibfnamefont {J.}~\bibnamefont {Stout}},\ and\ \bibinfo
  {author} {\bibfnamefont {G.~M.}\ \bibnamefont {Tomaselli}},\ }\bibfield
  {title} {\bibinfo {title} {{Ionization of gravitational atoms}},\ }\href
  {https://doi.org/10.1103/PhysRevD.105.115036} {\bibfield  {journal} {\bibinfo
   {journal} {Phys. Rev. D}\ }\textbf {\bibinfo {volume} {105}},\ \bibinfo
  {pages} {115036} (\bibinfo {year} {2022}{\natexlab{a}})},\ \Eprint
  {https://arxiv.org/abs/2112.14777} {arXiv:2112.14777 [gr-qc]} \BibitemShut
  {NoStop}%
\bibitem [{\citenamefont {Guo}\ \emph {et~al.}(2026{\natexlab{a}})\citenamefont
  {Guo}, \citenamefont {Zhong}, \citenamefont {Chen}, \citenamefont {Cardoso},
  \citenamefont {Ikeda},\ and\ \citenamefont {Zhou}}]{Guo:2025pea}%
  \BibitemOpen
  \bibfield  {author} {\bibinfo {author} {\bibfnamefont {Y.}~\bibnamefont
  {Guo}}, \bibinfo {author} {\bibfnamefont {Z.}~\bibnamefont {Zhong}}, \bibinfo
  {author} {\bibfnamefont {Y.}~\bibnamefont {Chen}}, \bibinfo {author}
  {\bibfnamefont {V.}~\bibnamefont {Cardoso}}, \bibinfo {author} {\bibfnamefont
  {T.}~\bibnamefont {Ikeda}},\ and\ \bibinfo {author} {\bibfnamefont
  {L.}~\bibnamefont {Zhou}},\ }\bibfield  {title} {\bibinfo {title}
  {{Ultralight boson ionization from comparable-mass binaries}},\ }\href
  {https://doi.org/10.1088/1475-7516/2026/08/065} {\bibfield  {journal}
  {\bibinfo  {journal} {JCAP}\ }\textbf {\bibinfo {volume} {08}},\ \bibinfo
  {pages} {065}},\ \Eprint {https://arxiv.org/abs/2509.09643} {arXiv:2509.09643
  [gr-qc]} \BibitemShut {NoStop}%
\bibitem [{\citenamefont {Guo}\ \emph {et~al.}(2023)\citenamefont {Guo},
  \citenamefont {Zhong}, \citenamefont {Ma},\ and\ \citenamefont
  {Su}}]{Guo:2023lbv}%
  \BibitemOpen
  \bibfield  {author} {\bibinfo {author} {\bibfnamefont {Y.}~\bibnamefont
  {Guo}}, \bibinfo {author} {\bibfnamefont {W.}~\bibnamefont {Zhong}}, \bibinfo
  {author} {\bibfnamefont {Y.}~\bibnamefont {Ma}},\ and\ \bibinfo {author}
  {\bibfnamefont {D.}~\bibnamefont {Su}},\ }\bibfield  {title} {\bibinfo
  {title} {{Mass transfer and boson cloud depletion in a binary black hole
  system}},\ }\href@noop {} {\  (\bibinfo {year} {2023})},\ \Eprint
  {https://arxiv.org/abs/2309.07790} {arXiv:2309.07790 [gr-qc]} \BibitemShut
  {NoStop}%
\bibitem [{\citenamefont {Guo}\ \emph {et~al.}(2024)\citenamefont {Guo},
  \citenamefont {Zhang},\ and\ \citenamefont {Yang}}]{Guo:2024iye}%
  \BibitemOpen
  \bibfield  {author} {\bibinfo {author} {\bibfnamefont {A.}~\bibnamefont
  {Guo}}, \bibinfo {author} {\bibfnamefont {J.}~\bibnamefont {Zhang}},\ and\
  \bibinfo {author} {\bibfnamefont {H.}~\bibnamefont {Yang}},\ }\bibfield
  {title} {\bibinfo {title} {{Superradiant clouds may be relevant for close
  compact object binaries}},\ }\href
  {https://doi.org/10.1103/PhysRevD.110.023022} {\bibfield  {journal} {\bibinfo
   {journal} {Phys. Rev. D}\ }\textbf {\bibinfo {volume} {110}},\ \bibinfo
  {pages} {023022} (\bibinfo {year} {2024})},\ \Eprint
  {https://arxiv.org/abs/2401.15003} {arXiv:2401.15003 [gr-qc]} \BibitemShut
  {NoStop}%
\bibitem [{\citenamefont {Guo}\ \emph {et~al.}(2026{\natexlab{b}})\citenamefont
  {Guo}, \citenamefont {Zhang}, \citenamefont {Yang},\ and\ \citenamefont
  {Zhang}}]{Guo:2025ckp}%
  \BibitemOpen
  \bibfield  {author} {\bibinfo {author} {\bibfnamefont {A.}~\bibnamefont
  {Guo}}, \bibinfo {author} {\bibfnamefont {Q.-Y.}\ \bibnamefont {Zhang}},
  \bibinfo {author} {\bibfnamefont {H.}~\bibnamefont {Yang}},\ and\ \bibinfo
  {author} {\bibfnamefont {J.}~\bibnamefont {Zhang}},\ }\bibfield  {title}
  {\bibinfo {title} {{Common envelope evolution of ultralight boson clouds}},\
  }\href {https://doi.org/10.1103/g79c-mqgx} {\bibfield  {journal} {\bibinfo
  {journal} {Phys. Rev. D}\ }\textbf {\bibinfo {volume} {113}},\ \bibinfo
  {pages} {043018} (\bibinfo {year} {2026}{\natexlab{b}})},\ \Eprint
  {https://arxiv.org/abs/2508.18738} {arXiv:2508.18738 [gr-qc]} \BibitemShut
  {NoStop}%
\bibitem [{\citenamefont {Zhang}\ \emph {et~al.}(2026)\citenamefont {Zhang},
  \citenamefont {Wei},\ and\ \citenamefont {Liu}}]{Zhang:2026buv}%
  \BibitemOpen
  \bibfield  {author} {\bibinfo {author} {\bibfnamefont {Y.-P.}\ \bibnamefont
  {Zhang}}, \bibinfo {author} {\bibfnamefont {S.-W.}\ \bibnamefont {Wei}},\
  and\ \bibinfo {author} {\bibfnamefont {Y.-X.}\ \bibnamefont {Liu}},\
  }\bibfield  {title} {\bibinfo {title} {{Scalar wave dark matter imprints on
  binary black hole mergers: dependence on field mass and density}},\
  }\href@noop {} {\  (\bibinfo {year} {2026})},\ \Eprint
  {https://arxiv.org/abs/2610.15053} {arXiv:2610.15053 [gr-qc]} \BibitemShut
  {NoStop}%
\bibitem [{\citenamefont {Cardoso}\ \emph {et~al.}(2011)\citenamefont
  {Cardoso}, \citenamefont {Chakrabarti}, \citenamefont {Pani}, \citenamefont
  {Berti},\ and\ \citenamefont {Gualtieri}}]{Cardoso:2011xi}%
  \BibitemOpen
  \bibfield  {author} {\bibinfo {author} {\bibfnamefont {V.}~\bibnamefont
  {Cardoso}}, \bibinfo {author} {\bibfnamefont {S.}~\bibnamefont
  {Chakrabarti}}, \bibinfo {author} {\bibfnamefont {P.}~\bibnamefont {Pani}},
  \bibinfo {author} {\bibfnamefont {E.}~\bibnamefont {Berti}},\ and\ \bibinfo
  {author} {\bibfnamefont {L.}~\bibnamefont {Gualtieri}},\ }\bibfield  {title}
  {\bibinfo {title} {{Floating and sinking: The Imprint of massive scalars
  around rotating black holes}},\ }\href
  {https://doi.org/10.1103/PhysRevLett.107.241101} {\bibfield  {journal}
  {\bibinfo  {journal} {Phys. Rev. Lett.}\ }\textbf {\bibinfo {volume} {107}},\
  \bibinfo {pages} {241101} (\bibinfo {year} {2011})},\ \Eprint
  {https://arxiv.org/abs/1109.6021} {arXiv:1109.6021 [gr-qc]} \BibitemShut
  {NoStop}%
\bibitem [{\citenamefont {Ferreira}\ \emph {et~al.}(2017)\citenamefont
  {Ferreira}, \citenamefont {Macedo},\ and\ \citenamefont
  {Cardoso}}]{Ferreira:2017pth}%
  \BibitemOpen
  \bibfield  {author} {\bibinfo {author} {\bibfnamefont {M.~C.}\ \bibnamefont
  {Ferreira}}, \bibinfo {author} {\bibfnamefont {C.~F.~B.}\ \bibnamefont
  {Macedo}},\ and\ \bibinfo {author} {\bibfnamefont {V.}~\bibnamefont
  {Cardoso}},\ }\bibfield  {title} {\bibinfo {title} {{Orbital fingerprints of
  ultralight scalar fields around black holes}},\ }\href
  {https://doi.org/10.1103/PhysRevD.96.083017} {\bibfield  {journal} {\bibinfo
  {journal} {Phys. Rev. D}\ }\textbf {\bibinfo {volume} {96}},\ \bibinfo
  {pages} {083017} (\bibinfo {year} {2017})},\ \Eprint
  {https://arxiv.org/abs/1710.00830} {arXiv:1710.00830 [gr-qc]} \BibitemShut
  {NoStop}%
\bibitem [{\citenamefont {Zhang}\ and\ \citenamefont
  {Yang}(2019)}]{Zhang:2018kib}%
  \BibitemOpen
  \bibfield  {author} {\bibinfo {author} {\bibfnamefont {J.}~\bibnamefont
  {Zhang}}\ and\ \bibinfo {author} {\bibfnamefont {H.}~\bibnamefont {Yang}},\
  }\bibfield  {title} {\bibinfo {title} {{Gravitational floating orbits around
  hairy black holes}},\ }\href {https://doi.org/10.1103/PhysRevD.99.064018}
  {\bibfield  {journal} {\bibinfo  {journal} {Phys. Rev. D}\ }\textbf {\bibinfo
  {volume} {99}},\ \bibinfo {pages} {064018} (\bibinfo {year} {2019})},\
  \Eprint {https://arxiv.org/abs/1808.02905} {arXiv:1808.02905 [gr-qc]}
  \BibitemShut {NoStop}%
\bibitem [{\citenamefont {Zhang}\ and\ \citenamefont
  {Yang}(2020)}]{Zhang:2019eid}%
  \BibitemOpen
  \bibfield  {author} {\bibinfo {author} {\bibfnamefont {J.}~\bibnamefont
  {Zhang}}\ and\ \bibinfo {author} {\bibfnamefont {H.}~\bibnamefont {Yang}},\
  }\bibfield  {title} {\bibinfo {title} {{Dynamic Signatures of Black Hole
  Binaries with Superradiant Clouds}},\ }\href
  {https://doi.org/10.1103/PhysRevD.101.043020} {\bibfield  {journal} {\bibinfo
   {journal} {Phys. Rev. D}\ }\textbf {\bibinfo {volume} {101}},\ \bibinfo
  {pages} {043020} (\bibinfo {year} {2020})},\ \Eprint
  {https://arxiv.org/abs/1907.13582} {arXiv:1907.13582 [gr-qc]} \BibitemShut
  {NoStop}%
\bibitem [{\citenamefont {Baumann}\ \emph
  {et~al.}(2022{\natexlab{b}})\citenamefont {Baumann}, \citenamefont {Bertone},
  \citenamefont {Stout},\ and\ \citenamefont {Tomaselli}}]{Baumann:2022pkl}%
  \BibitemOpen
  \bibfield  {author} {\bibinfo {author} {\bibfnamefont {D.}~\bibnamefont
  {Baumann}}, \bibinfo {author} {\bibfnamefont {G.}~\bibnamefont {Bertone}},
  \bibinfo {author} {\bibfnamefont {J.}~\bibnamefont {Stout}},\ and\ \bibinfo
  {author} {\bibfnamefont {G.~M.}\ \bibnamefont {Tomaselli}},\ }\bibfield
  {title} {\bibinfo {title} {{Sharp Signals of Boson Clouds in Black Hole
  Binary Inspirals}},\ }\href {https://doi.org/10.1103/PhysRevLett.128.221102}
  {\bibfield  {journal} {\bibinfo  {journal} {Phys. Rev. Lett.}\ }\textbf
  {\bibinfo {volume} {128}},\ \bibinfo {pages} {221102} (\bibinfo {year}
  {2022}{\natexlab{b}})},\ \Eprint {https://arxiv.org/abs/2206.01212}
  {arXiv:2206.01212 [gr-qc]} \BibitemShut {NoStop}%
\bibitem [{\citenamefont {Tomaselli}\ \emph {et~al.}(2023)\citenamefont
  {Tomaselli}, \citenamefont {Spieksma},\ and\ \citenamefont
  {Bertone}}]{Tomaselli:2023ysb}%
  \BibitemOpen
  \bibfield  {author} {\bibinfo {author} {\bibfnamefont {G.~M.}\ \bibnamefont
  {Tomaselli}}, \bibinfo {author} {\bibfnamefont {T.~F.~M.}\ \bibnamefont
  {Spieksma}},\ and\ \bibinfo {author} {\bibfnamefont {G.}~\bibnamefont
  {Bertone}},\ }\bibfield  {title} {\bibinfo {title} {{Dynamical friction in
  gravitational atoms}},\ }\href
  {https://doi.org/10.1088/1475-7516/2023/07/070} {\bibfield  {journal}
  {\bibinfo  {journal} {JCAP}\ }\textbf {\bibinfo {volume} {07}},\ \bibinfo
  {pages} {070}},\ \Eprint {https://arxiv.org/abs/2305.15460} {arXiv:2305.15460
  [gr-qc]} \BibitemShut {NoStop}%
\bibitem [{\citenamefont {Cao}\ and\ \citenamefont {Tang}(2023)}]{Cao:2023fyv}%
  \BibitemOpen
  \bibfield  {author} {\bibinfo {author} {\bibfnamefont {Y.}~\bibnamefont
  {Cao}}\ and\ \bibinfo {author} {\bibfnamefont {Y.}~\bibnamefont {Tang}},\
  }\bibfield  {title} {\bibinfo {title} {{Signatures of Ultralight Bosons in
  Compact Binary Inspiral and Outspiral}},\ }\href@noop {} {\  (\bibinfo {year}
  {2023})},\ \Eprint {https://arxiv.org/abs/2307.05181} {arXiv:2307.05181
  [gr-qc]} \BibitemShut {NoStop}%
\bibitem [{\citenamefont {Tomaselli}\ \emph
  {et~al.}(2024{\natexlab{a}})\citenamefont {Tomaselli}, \citenamefont
  {Spieksma},\ and\ \citenamefont {Bertone}}]{Tomaselli:2024bdd}%
  \BibitemOpen
  \bibfield  {author} {\bibinfo {author} {\bibfnamefont {G.~M.}\ \bibnamefont
  {Tomaselli}}, \bibinfo {author} {\bibfnamefont {T.~F.~M.}\ \bibnamefont
  {Spieksma}},\ and\ \bibinfo {author} {\bibfnamefont {G.}~\bibnamefont
  {Bertone}},\ }\bibfield  {title} {\bibinfo {title} {{Resonant history of
  gravitational atoms in black hole binaries}},\ }\href
  {https://doi.org/10.1103/PhysRevD.110.064048} {\bibfield  {journal} {\bibinfo
   {journal} {Phys. Rev. D}\ }\textbf {\bibinfo {volume} {110}},\ \bibinfo
  {pages} {064048} (\bibinfo {year} {2024}{\natexlab{a}})},\ \Eprint
  {https://arxiv.org/abs/2403.03147} {arXiv:2403.03147 [gr-qc]} \BibitemShut
  {NoStop}%
\bibitem [{\citenamefont {Cao}\ \emph {et~al.}(2024)\citenamefont {Cao},
  \citenamefont {Cheng}, \citenamefont {Li},\ and\ \citenamefont
  {Tang}}]{Cao:2024wby}%
  \BibitemOpen
  \bibfield  {author} {\bibinfo {author} {\bibfnamefont {Y.}~\bibnamefont
  {Cao}}, \bibinfo {author} {\bibfnamefont {Y.-Z.}\ \bibnamefont {Cheng}},
  \bibinfo {author} {\bibfnamefont {G.-L.}\ \bibnamefont {Li}},\ and\ \bibinfo
  {author} {\bibfnamefont {Y.}~\bibnamefont {Tang}},\ }\bibfield  {title}
  {\bibinfo {title} {{Probing vector gravitational atom with eccentric
  intermediate mass-ratio inspirals}},\ }\href@noop {} {\  (\bibinfo {year}
  {2024})},\ \Eprint {https://arxiv.org/abs/2411.17247} {arXiv:2411.17247
  [gr-qc]} \BibitemShut {NoStop}%
\bibitem [{\citenamefont {Takahashi}\ \emph {et~al.}(2024)\citenamefont
  {Takahashi}, \citenamefont {Omiya},\ and\ \citenamefont
  {Tanaka}}]{Takahashi:2024fyq}%
  \BibitemOpen
  \bibfield  {author} {\bibinfo {author} {\bibfnamefont {T.}~\bibnamefont
  {Takahashi}}, \bibinfo {author} {\bibfnamefont {H.}~\bibnamefont {Omiya}},\
  and\ \bibinfo {author} {\bibfnamefont {T.}~\bibnamefont {Tanaka}},\
  }\bibfield  {title} {\bibinfo {title} {{Self-interacting axion clouds around
  rotating black holes in binary systems}},\ }\href
  {https://doi.org/10.1103/PhysRevD.110.104038} {\bibfield  {journal} {\bibinfo
   {journal} {Phys. Rev. D}\ }\textbf {\bibinfo {volume} {110}},\ \bibinfo
  {pages} {104038} (\bibinfo {year} {2024})},\ \Eprint
  {https://arxiv.org/abs/2408.08349} {arXiv:2408.08349 [gr-qc]} \BibitemShut
  {NoStop}%
\bibitem [{\citenamefont {Zhu}\ \emph {et~al.}(2025)\citenamefont {Zhu},
  \citenamefont {Tong}, \citenamefont {Manzoni},\ and\ \citenamefont
  {Ma}}]{Zhu:2024bqs}%
  \BibitemOpen
  \bibfield  {author} {\bibinfo {author} {\bibfnamefont {H.-Y.}\ \bibnamefont
  {Zhu}}, \bibinfo {author} {\bibfnamefont {X.}~\bibnamefont {Tong}}, \bibinfo
  {author} {\bibfnamefont {G.}~\bibnamefont {Manzoni}},\ and\ \bibinfo {author}
  {\bibfnamefont {Y.}~\bibnamefont {Ma}},\ }\bibfield  {title} {\bibinfo
  {title} {{Survival of the Fittest: Testing Superradiance Termination with
  Simulated Binary Black Hole Statistics}},\ }\href
  {https://doi.org/10.3847/1538-4357/adb1db} {\bibfield  {journal} {\bibinfo
  {journal} {Astrophys. J.}\ }\textbf {\bibinfo {volume} {981}},\ \bibinfo
  {pages} {165} (\bibinfo {year} {2025})},\ \Eprint
  {https://arxiv.org/abs/2409.14159} {arXiv:2409.14159 [gr-qc]} \BibitemShut
  {NoStop}%
\bibitem [{\citenamefont {Tomaselli}\ \emph
  {et~al.}(2024{\natexlab{b}})\citenamefont {Tomaselli}, \citenamefont
  {Spieksma},\ and\ \citenamefont {Bertone}}]{Tomaselli:2024dbw}%
  \BibitemOpen
  \bibfield  {author} {\bibinfo {author} {\bibfnamefont {G.~M.}\ \bibnamefont
  {Tomaselli}}, \bibinfo {author} {\bibfnamefont {T.~F.~M.}\ \bibnamefont
  {Spieksma}},\ and\ \bibinfo {author} {\bibfnamefont {G.}~\bibnamefont
  {Bertone}},\ }\bibfield  {title} {\bibinfo {title} {{Legacy of Boson Clouds
  on Black Hole Binaries}},\ }\href
  {https://doi.org/10.1103/PhysRevLett.133.121402} {\bibfield  {journal}
  {\bibinfo  {journal} {Phys. Rev. Lett.}\ }\textbf {\bibinfo {volume} {133}},\
  \bibinfo {pages} {121402} (\bibinfo {year} {2024}{\natexlab{b}})},\ \Eprint
  {https://arxiv.org/abs/2407.12908} {arXiv:2407.12908 [gr-qc]} \BibitemShut
  {NoStop}%
\bibitem [{\citenamefont {Bo\v{s}kovi\'c}\ \emph {et~al.}(2024)\citenamefont
  {Bo\v{s}kovi\'c}, \citenamefont {Koschnitzke},\ and\ \citenamefont
  {Porto}}]{Boskovic:2024fga}%
  \BibitemOpen
  \bibfield  {author} {\bibinfo {author} {\bibfnamefont {M.}~\bibnamefont
  {Bo\v{s}kovi\'c}}, \bibinfo {author} {\bibfnamefont {M.}~\bibnamefont
  {Koschnitzke}},\ and\ \bibinfo {author} {\bibfnamefont {R.~A.}\ \bibnamefont
  {Porto}},\ }\bibfield  {title} {\bibinfo {title} {{Signatures of Ultralight
  Bosons in the Orbital Eccentricity of Binary Black Holes}},\ }\href
  {https://doi.org/10.1103/PhysRevLett.133.121401} {\bibfield  {journal}
  {\bibinfo  {journal} {Phys. Rev. Lett.}\ }\textbf {\bibinfo {volume} {133}},\
  \bibinfo {pages} {121401} (\bibinfo {year} {2024})},\ \Eprint
  {https://arxiv.org/abs/2403.02415} {arXiv:2403.02415 [gr-qc]} \BibitemShut
  {NoStop}%
\bibitem [{\citenamefont {Kyriazis}\ and\ \citenamefont
  {Yang}(2025)}]{Kyriazis:2025fis}%
  \BibitemOpen
  \bibfield  {author} {\bibinfo {author} {\bibfnamefont {A.}~\bibnamefont
  {Kyriazis}}\ and\ \bibinfo {author} {\bibfnamefont {F.}~\bibnamefont
  {Yang}},\ }\bibfield  {title} {\bibinfo {title} {{Gravitational Waves from
  Resonant Transitions of Tidally Perturbed Gravitational Atoms}},\ }\href@noop
  {} {\  (\bibinfo {year} {2025})},\ \Eprint {https://arxiv.org/abs/2503.18121}
  {arXiv:2503.18121 [hep-ph]} \BibitemShut {NoStop}%
\bibitem [{\citenamefont {Peng}\ and\ \citenamefont
  {Zhang}(2026)}]{Peng:2025zca}%
  \BibitemOpen
  \bibfield  {author} {\bibinfo {author} {\bibfnamefont {S.-T.}\ \bibnamefont
  {Peng}}\ and\ \bibinfo {author} {\bibfnamefont {J.}~\bibnamefont {Zhang}},\
  }\bibfield  {title} {\bibinfo {title} {{Gravitational waves from superradiant
  cloud level transition}},\ }\href {https://doi.org/10.1103/s69g-1wyq}
  {\bibfield  {journal} {\bibinfo  {journal} {Phys. Rev. D}\ }\textbf {\bibinfo
  {volume} {113}},\ \bibinfo {pages} {064038} (\bibinfo {year} {2026})},\
  \Eprint {https://arxiv.org/abs/2504.00728} {arXiv:2504.00728 [gr-qc]}
  \BibitemShut {NoStop}%
\bibitem [{\citenamefont {Roy}\ \emph {et~al.}(2026)\citenamefont {Roy},
  \citenamefont {Vicente}, \citenamefont {Aurrekoetxea}, \citenamefont
  {Clough},\ and\ \citenamefont {Ferreira}}]{Roy:2025qaa}%
  \BibitemOpen
  \bibfield  {author} {\bibinfo {author} {\bibfnamefont {S.}~\bibnamefont
  {Roy}}, \bibinfo {author} {\bibfnamefont {R.}~\bibnamefont {Vicente}},
  \bibinfo {author} {\bibfnamefont {J.~C.}\ \bibnamefont {Aurrekoetxea}},
  \bibinfo {author} {\bibfnamefont {K.}~\bibnamefont {Clough}},\ and\ \bibinfo
  {author} {\bibfnamefont {P.~G.}\ \bibnamefont {Ferreira}},\ }\bibfield
  {title} {\bibinfo {title} {{Scalar Fields around Black Hole Binaries in
  LIGO-Virgo-KAGRA}},\ }\href {https://doi.org/10.1103/fv9z-zkxx} {\bibfield
  {journal} {\bibinfo  {journal} {Phys. Rev. Lett.}\ }\textbf {\bibinfo
  {volume} {136}},\ \bibinfo {pages} {191402} (\bibinfo {year} {2026})},\
  \Eprint {https://arxiv.org/abs/2510.17967} {arXiv:2510.17967 [gr-qc]}
  \BibitemShut {NoStop}%
\bibitem [{\citenamefont {Chen}\ \emph {et~al.}(2024)\citenamefont {Chen},
  \citenamefont {Liu}, \citenamefont {Zhang},\ and\ \citenamefont
  {Zhang}}]{Chen:2024ery}%
  \BibitemOpen
  \bibfield  {author} {\bibinfo {author} {\bibfnamefont {M.-C.}\ \bibnamefont
  {Chen}}, \bibinfo {author} {\bibfnamefont {H.-Y.}\ \bibnamefont {Liu}},
  \bibinfo {author} {\bibfnamefont {Q.-Y.}\ \bibnamefont {Zhang}},\ and\
  \bibinfo {author} {\bibfnamefont {J.}~\bibnamefont {Zhang}},\ }\bibfield
  {title} {\bibinfo {title} {{Probing massive fields with multiband
  gravitational-wave observations}},\ }\href
  {https://doi.org/10.1103/PhysRevD.110.064018} {\bibfield  {journal} {\bibinfo
   {journal} {Phys. Rev. D}\ }\textbf {\bibinfo {volume} {110}},\ \bibinfo
  {pages} {064018} (\bibinfo {year} {2024})},\ \Eprint
  {https://arxiv.org/abs/2405.11583} {arXiv:2405.11583 [gr-qc]} \BibitemShut
  {NoStop}%
\bibitem [{\citenamefont {Abac}\ \emph {et~al.}(2025)\citenamefont {Abac} \emph
  {et~al.}}]{LIGOScientific:2025jau}%
  \BibitemOpen
  \bibfield  {author} {\bibinfo {author} {\bibfnamefont {A.~G.}\ \bibnamefont
  {Abac}} \emph {et~al.} (\bibinfo {collaboration} {LIGO Scientific, VIRGO,
  KAGRA}),\ }\bibfield  {title} {\bibinfo {title} {{GWTC-4.0: Constraints on
  the Cosmic Expansion Rate and Modified Gravitational-wave Propagation}},\
  }\href@noop {} {\bibfield  {journal} {\bibinfo  {journal} {arXiv:2509.04348}\
  } (\bibinfo {year} {2025})},\ \Eprint {https://arxiv.org/abs/2509.04348}
  {arXiv:2509.04348 [astro-ph.CO]} \BibitemShut {NoStop}%
\bibitem [{\citenamefont {Madau}\ and\ \citenamefont
  {Dickinson}(2014)}]{Madau:2014bja}%
  \BibitemOpen
  \bibfield  {author} {\bibinfo {author} {\bibfnamefont {P.}~\bibnamefont
  {Madau}}\ and\ \bibinfo {author} {\bibfnamefont {M.}~\bibnamefont
  {Dickinson}},\ }\bibfield  {title} {\bibinfo {title} {{Cosmic Star Formation
  History}},\ }\href {https://doi.org/10.1146/annurev-astro-081811-125615}
  {\bibfield  {journal} {\bibinfo  {journal} {Ann. Rev. Astron. Astrophys.}\
  }\textbf {\bibinfo {volume} {52}},\ \bibinfo {pages} {415} (\bibinfo {year}
  {2014})},\ \Eprint {https://arxiv.org/abs/1403.0007} {arXiv:1403.0007
  [astro-ph.CO]} \BibitemShut {NoStop}%
\bibitem [{\citenamefont {Aghanim}\ \emph {et~al.}(2020)\citenamefont {Aghanim}
  \emph {et~al.}}]{Planck:2018vyg}%
  \BibitemOpen
  \bibfield  {author} {\bibinfo {author} {\bibfnamefont {N.}~\bibnamefont
  {Aghanim}} \emph {et~al.} (\bibinfo {collaboration} {Planck}),\ }\bibfield
  {title} {\bibinfo {title} {{Planck 2018 results. VI. Cosmological
  parameters}},\ }\href {https://doi.org/10.1051/0004-6361/201833910}
  {\bibfield  {journal} {\bibinfo  {journal} {Astron. Astrophys.}\ }\textbf
  {\bibinfo {volume} {641}},\ \bibinfo {pages} {A6} (\bibinfo {year} {2020})},\
  \bibinfo {note} {[Erratum: Astron.Astrophys. 652, C4 (2021)]},\ \Eprint
  {https://arxiv.org/abs/1807.06209} {arXiv:1807.06209 [astro-ph.CO]}
  \BibitemShut {NoStop}%
\bibitem [{\citenamefont {Gerosa}\ \emph {et~al.}(2019)\citenamefont {Gerosa},
  \citenamefont {Ma}, \citenamefont {Wong}, \citenamefont {Berti},
  \citenamefont {O'Shaughnessy}, \citenamefont {Chen},\ and\ \citenamefont
  {Belczynski}}]{Gerosa:2019dbe}%
  \BibitemOpen
  \bibfield  {author} {\bibinfo {author} {\bibfnamefont {D.}~\bibnamefont
  {Gerosa}}, \bibinfo {author} {\bibfnamefont {S.}~\bibnamefont {Ma}}, \bibinfo
  {author} {\bibfnamefont {K.~W.~K.}\ \bibnamefont {Wong}}, \bibinfo {author}
  {\bibfnamefont {E.}~\bibnamefont {Berti}}, \bibinfo {author} {\bibfnamefont
  {R.}~\bibnamefont {O'Shaughnessy}}, \bibinfo {author} {\bibfnamefont
  {Y.}~\bibnamefont {Chen}},\ and\ \bibinfo {author} {\bibfnamefont
  {K.}~\bibnamefont {Belczynski}},\ }\bibfield  {title} {\bibinfo {title}
  {{Multiband gravitational-wave event rates and stellar physics}},\ }\href
  {https://doi.org/10.1103/PhysRevD.99.103004} {\bibfield  {journal} {\bibinfo
  {journal} {Phys. Rev. D}\ }\textbf {\bibinfo {volume} {99}},\ \bibinfo
  {pages} {103004} (\bibinfo {year} {2019})},\ \Eprint
  {https://arxiv.org/abs/1902.00021} {arXiv:1902.00021 [astro-ph.HE]}
  \BibitemShut {NoStop}%
\bibitem [{\citenamefont {Ajith}\ \emph {et~al.}(2008)\citenamefont {Ajith}
  \emph {et~al.}}]{Ajith:2007kx}%
  \BibitemOpen
  \bibfield  {author} {\bibinfo {author} {\bibfnamefont {P.}~\bibnamefont
  {Ajith}} \emph {et~al.},\ }\bibfield  {title} {\bibinfo {title} {{A Template
  bank for gravitational waveforms from coalescing binary black holes. I.
  Non-spinning binaries}},\ }\href {https://doi.org/10.1103/PhysRevD.77.104017}
  {\bibfield  {journal} {\bibinfo  {journal} {Phys. Rev. D}\ }\textbf {\bibinfo
  {volume} {77}},\ \bibinfo {pages} {104017} (\bibinfo {year} {2008})},\
  \bibinfo {note} {[Erratum: Phys.Rev.D 79, 129901 (2009)]},\ \Eprint
  {https://arxiv.org/abs/0710.2335} {arXiv:0710.2335 [gr-qc]} \BibitemShut
  {NoStop}%
\bibitem [{\citenamefont {Robson}\ \emph {et~al.}(2019)\citenamefont {Robson},
  \citenamefont {Cornish},\ and\ \citenamefont {Liu}}]{Robson:2018ifk}%
  \BibitemOpen
  \bibfield  {author} {\bibinfo {author} {\bibfnamefont {T.}~\bibnamefont
  {Robson}}, \bibinfo {author} {\bibfnamefont {N.~J.}\ \bibnamefont
  {Cornish}},\ and\ \bibinfo {author} {\bibfnamefont {C.}~\bibnamefont {Liu}},\
  }\bibfield  {title} {\bibinfo {title} {{The construction and use of LISA
  sensitivity curves}},\ }\href {https://doi.org/10.1088/1361-6382/ab1101}
  {\bibfield  {journal} {\bibinfo  {journal} {Class. Quant. Grav.}\ }\textbf
  {\bibinfo {volume} {36}},\ \bibinfo {pages} {105011} (\bibinfo {year}
  {2019})},\ \Eprint {https://arxiv.org/abs/1803.01944} {arXiv:1803.01944
  [astro-ph.HE]} \BibitemShut {NoStop}%
\bibitem [{\citenamefont {Abac}\ \emph {et~al.}(2026)\citenamefont {Abac} \emph
  {et~al.}}]{LIGOScientific:2026vua}%
  \BibitemOpen
  \bibfield  {author} {\bibinfo {author} {\bibfnamefont {A.~G.}\ \bibnamefont
  {Abac}} \emph {et~al.} (\bibinfo {collaboration} {LIGO Scientific, Virgo,,
  KAGRA}),\ }\bibfield  {title} {\bibinfo {title} {{GWTC-4.0: Population
  Properties of Merging Compact Binaries}},\ }\href
  {https://doi.org/10.3847/2041-8213/ae771e} {\bibfield  {journal} {\bibinfo
  {journal} {Astrophys. J. Lett.}\ }\textbf {\bibinfo {volume} {1005}},\
  \bibinfo {pages} {L51} (\bibinfo {year} {2026})}\BibitemShut {NoStop}%
\bibitem [{\citenamefont {Buscicchio}\ \emph {et~al.}(2025)\citenamefont
  {Buscicchio}, \citenamefont {Torrado}, \citenamefont {Caprini}, \citenamefont
  {Nardini}, \citenamefont {Karnesis}, \citenamefont {Pieroni},\ and\
  \citenamefont {Sesana}}]{Buscicchio:2024asl}%
  \BibitemOpen
  \bibfield  {author} {\bibinfo {author} {\bibfnamefont {R.}~\bibnamefont
  {Buscicchio}}, \bibinfo {author} {\bibfnamefont {J.}~\bibnamefont {Torrado}},
  \bibinfo {author} {\bibfnamefont {C.}~\bibnamefont {Caprini}}, \bibinfo
  {author} {\bibfnamefont {G.}~\bibnamefont {Nardini}}, \bibinfo {author}
  {\bibfnamefont {N.}~\bibnamefont {Karnesis}}, \bibinfo {author}
  {\bibfnamefont {M.}~\bibnamefont {Pieroni}},\ and\ \bibinfo {author}
  {\bibfnamefont {A.}~\bibnamefont {Sesana}},\ }\bibfield  {title} {\bibinfo
  {title} {{Stellar-mass black-hole binaries in LISA: characteristics and
  complementarity with current-generation interferometers}},\ }\href
  {https://doi.org/10.1088/1475-7516/2025/01/084} {\bibfield  {journal}
  {\bibinfo  {journal} {JCAP}\ }\textbf {\bibinfo {volume} {2025}}\bibfield
  {number} {\bibinfo  {number} { (01)},\ \bibinfo {pages} {084}},\ }\Eprint
  {https://arxiv.org/abs/2410.18171} {arXiv:2410.18171 [astro-ph.HE]}
  \BibitemShut {NoStop}%
\bibitem [{\citenamefont {Wong}\ \emph {et~al.}(2018)\citenamefont {Wong},
  \citenamefont {Kovetz}, \citenamefont {Cutler},\ and\ \citenamefont
  {Berti}}]{Wong:2018uwb}%
  \BibitemOpen
  \bibfield  {author} {\bibinfo {author} {\bibfnamefont {K.~W.~K.}\
  \bibnamefont {Wong}}, \bibinfo {author} {\bibfnamefont {E.~D.}\ \bibnamefont
  {Kovetz}}, \bibinfo {author} {\bibfnamefont {C.}~\bibnamefont {Cutler}},\
  and\ \bibinfo {author} {\bibfnamefont {E.}~\bibnamefont {Berti}},\ }\bibfield
   {title} {\bibinfo {title} {{Expanding the LISA Horizon from the Ground}},\
  }\href {https://doi.org/10.1103/PhysRevLett.121.251102} {\bibfield  {journal}
  {\bibinfo  {journal} {Phys. Rev. Lett.}\ }\textbf {\bibinfo {volume} {121}},\
  \bibinfo {pages} {251102} (\bibinfo {year} {2018})},\ \Eprint
  {https://arxiv.org/abs/1808.08247} {arXiv:1808.08247 [astro-ph.HE]}
  \BibitemShut {NoStop}%
\bibitem [{\citenamefont {Ashton}\ \emph {et~al.}(2019)\citenamefont {Ashton}
  \emph {et~al.}}]{bilby_paper}%
  \BibitemOpen
  \bibfield  {author} {\bibinfo {author} {\bibfnamefont {G.}~\bibnamefont
  {Ashton}} \emph {et~al.},\ }\bibfield  {title} {\bibinfo {title} {{BILBY: A
  user-friendly Bayesian inference library for gravitational-wave astronomy}},\
  }\href {https://doi.org/10.3847/1538-4365/ab06fc} {\bibfield  {journal}
  {\bibinfo  {journal} {Astrophys. J. Suppl.}\ }\textbf {\bibinfo {volume}
  {241}},\ \bibinfo {pages} {27} (\bibinfo {year} {2019})},\ \Eprint
  {https://arxiv.org/abs/1811.02042} {arXiv:1811.02042 [astro-ph.IM]}
  \BibitemShut {NoStop}%
\end{thebibliography}%

\end{document}